\documentclass[a4paper,UKenglish,cleveref, autoref, thm-restate]{lipics-v2021}

\title{Linear Time Runs over\\General Ordered Alphabets\footnote{This work has been submitted to ICALP 2021.}} 

\titlerunning{Linear Time Runs over General Ordered Alphabets} 

\author{Jonas Ellert}{Department of Computer Science,\\Techincal University of Dortmund, Germany}{jonas.ellert@tu-dortmund.de}{https://orcid.org/0000-0003-3305-6185}{}

\author{Johannes Fischer}{Department of Computer Science,\\Techincal University of Dortmund, Germany}{johannes.fischer@cs.tu-dortmund.de}{}{}

\authorrunning{J. Ellert and J. Fischer} 

\Copyright{Jonas Ellert and Johannes Fischer} 



\category{} 

\relatedversion{} 

\supplement{\url{https://github.com/jonas-ellert/linear-time-runs/}}

\nolinenumbers 

\hideLIPIcs  

\EventEditors{John Q. Open and Joan R. Access}
\EventNoEds{2}
\EventLongTitle{42nd Conference on Very Important Topics (CVIT 2016)}
\EventShortTitle{CVIT 2016}
\EventAcronym{CVIT}
\EventYear{2016}
\EventDate{December 24--27, 2016}
\EventLocation{Little Whinging, United Kingdom}
\EventLogo{}
\SeriesVolume{42}
\ArticleNo{23}

\usepackage{amsmath}
\usepackage{graphicx}
\usepackage{url}
\usepackage{cite}
\usepackage{mathtools}
\usepackage[capitalise]{cleveref}
\usepackage{multirow}
\usepackage{verbatim}
\usepackage{amsfonts}
\usepackage{placeins}
\usepackage{tabularx}
\usepackage{amssymb}
\usepackage{refcount}
\usepackage{needspace}
\usepackage[skins,breakable]{tcolorbox}
\usepackage{etoolbox}
\usepackage[shortcuts]{extdash}

\newcommand{\lipicshl}[1]{\textcolor{lipicsGray}{\sffamily\bfseries\upshape\mathversion{bold}#1}}
\newcommand{\case}[1]{\lipicshl{Case #1:}}

\usepackage{algorithm}
\usepackage[noend]{algpseudocode}

\errorcontextlines\maxdimen

\newcommand{\ALGtikzmarkcolor}{black}
\newcommand{\ALGtikzmarkextraindent}{4pt}
\newcommand{\ALGtikzmarkverticaloffsetstart}{-.5ex}
\newcommand{\ALGtikzmarkverticaloffsetend}{-.5ex}
\makeatletter
\newcounter{ALG@tikzmark@tempcnta}

\newcommand\ALG@tikzmark@start{%
    \global\let\ALG@tikzmark@last\ALG@tikzmark@starttext%
    \expandafter\edef\csname ALG@tikzmark@\theALG@nested\endcsname{\theALG@tikzmark@tempcnta}%
    \tikzmark{ALG@tikzmark@start@\csname ALG@tikzmark@\theALG@nested\endcsname}%
    \addtocounter{ALG@tikzmark@tempcnta}{1}%
}

\def\ALG@tikzmark@starttext{start}
\newcommand\ALG@tikzmark@end{%
    \ifx\ALG@tikzmark@last\ALG@tikzmark@starttext
    \else
    \tikzmark{ALG@tikzmark@end@\csname ALG@tikzmark@\theALG@nested\endcsname}%
    \tikz[overlay,remember picture] \draw[\ALGtikzmarkcolor] let \p{S}=($(ALG@tikzmark@start@\csname ALG@tikzmark@\theALG@nested\endcsname)+(\ALGtikzmarkextraindent,\ALGtikzmarkverticaloffsetstart)$), \p{E}=($(ALG@tikzmark@end@\csname ALG@tikzmark@\theALG@nested\endcsname)+(\ALGtikzmarkextraindent,\ALGtikzmarkverticaloffsetend)$) in (\x{S},\y{S})--(\x{S},\y{E});%
    \fi
    \gdef\ALG@tikzmark@last{end}%
}

\apptocmd{\ALG@beginblock}{\ALG@tikzmark@start}{}{\errmessage{failed to patch}}
\pretocmd{\ALG@endblock}{\ALG@tikzmark@end}{}{\errmessage{failed to patch}}
\makeatother

\definecolor{my-third}{HTML}{7570b3}
\definecolor{my-first}{HTML}{1b9e77}
\definecolor{my-second}{HTML}{d95f02}

\usetikzlibrary{positioning, calc, arrows, arrows.meta}
\newcommand{\tikzmark}[1]{\begin{tikzpicture}[overlay, remember picture]
  \node[inner sep=0pt, outer sep=0pt] (#1) {};
\end{tikzpicture}}

\newcommand{\orderof}[1]{\ensuremath{\mathcal{O}(#1)}}

\newcommand{\tocite}[2][]{\def\temp{#1}\ifx\temp\empty
  \cite{TODO}\todo{cite #2}%
\else
  \cite[#1]{TODO}\todo{cite #2}%
\fi}

\newcommand{\toref}[1]{\cref{TODO}\todo{ref #1}}

\newcommand{\ceil}[1]{\left\lceil#1\right\rceil}
\newcommand{\absolute}[1]{\left\lvert#1\right\rvert}
\newcommand{\llex}{\prec}

\newcommand{\leqlex}{\preceq}

\newcommand{\glex}{\succ}

\newcommand{\geqlex}{\succeq}

\newcommand{\str}{S}
\newcommand{\stralt}{T}
\newcommand{\nss}[1]{\textnormal{\textsf{nss}}[#1]}

\newcommand{\lyndarr}[1]{\lambda[#1]}

\newcommand{\examplestr}[1]{\texttt{#1}}
\newcommand{\triple}[1]{\langle#1\rangle}
\newcommand{\lcp}[1]{\textsc{lcp}(#1)}
\newcommand{\lcs}[1]{\textsc{lcs}(#1)}
\newcommand{\bm}[1]{\text{\boldmath$#1$\unboldmath}}

\makeatletter
\newcommand{\overrightharpoon}{%
  \mathpalette{\overarrow@\rightharpoonfill@}}
\def\rightharpoonfill@{\arrowfill@\relbar\relbar\rightharpoonup}
\newcommand{\overleftharpoon}{%
  \mathpalette{\overarrow@\leftharpoonfill@}}
\def\leftharpoonfill@{\arrowfill@\leftharpoonup\relbar\relbar}
\makeatother

\newcommand{\rlce}[1]{\textsc{lce}_{r}(#1)}
\newcommand{\llce}[1]{\textsc{lce}_{\ell}(#1)}
\newcommand{\ellmax}{\ell_{\max}}
\newcommand{\xhs}[1]{{%
\overset{#1}{\smash{c}\vphantom{\raisebox{1.15mm}{}}}
}}
\newcommand{\rhs}{\xhs{\rightarrow}}
\newcommand{\lhs}{\xhs{\leftarrow}}

\newlength{\strboxsep}
\newlength{\strboxrule}
\newlength{\strboxspace}
\newcommand{\stringbox}[4][.7em]{%
\renewcommand{\fboxsep}{\strboxsep}%
\renewcommand{\fboxrule}{\strboxrule}%
\fbox{$\vphantom{f}$\smash{\rlap{$#2$}\hspace{#1}$#3$\hspace{#1}\llap{$#4$}}}}

\newcommand{\stringidx}[2][\phantom{.}]{\overset{\mathclap{\underset{\scriptstyle\downarrow}{\scriptstyle\vphantom{(ji\rhs\lhs-+}#2}}}{\vphantom{\stringbox{}{}{}}#1}}
\newcommand{\stringstrut}{\vphantom{\stringbox{}{\stringidx{\ell}}{}}}

\usepackage{microtype}

\begin{document}

\maketitle

\begin{abstract}
A run in a string is a maximal periodic substring. For example, the string \texttt{bananatree} contains the runs $\texttt{anana} = (\texttt{an})^{3/2}$ and $\texttt{ee} = \texttt{e}^2$.
There are less than $n$ runs in any length-$n$ string, and computing all runs for a string over a linearly-sortable alphabet takes $\orderof{n}$ time (Bannai et al., SODA 2015). Kosolobov conjectured that there also exists a linear time runs algorithm for general ordered alphabets (Inf. Process. Lett. 2016). The conjecture was almost proven by Crochemore et al., who presented an $\orderof{n\alpha(n)}$ time algorithm (where $\alpha(n)$ is the extremely slowly growing inverse Ackermann function). We show how to achieve $\orderof{n}$ time by exploiting combinatorial properties of the Lyndon array, thus proving Kosolobov's conjecture.
\end{abstract}

\section{Introduction and Related Work}

A run in a string $\str$ is a maximal periodic substring. For example, the string $\str = \examplestr{bananatree}$ contains exactly the runs $\examplestr{anana} = (\examplestr{an})^{3/2}$ and $\examplestr{ee} = \examplestr{e}^2$. 
Identifying such repetitive structures in strings is of great importance for applications like text compression, text indexing and computational biology (for a general overview see \cite{Crochemore2009}).
To name just one example, runs in human genes (called maximal tandem repeats) are involved with a number of neurological disorders \cite{Budworth2013}.
In 1999, Kolpakov and Kucherov showed that the maximum number $\rho(n)$ of runs in a length-$n$ string is bounded by $\orderof{n}$, and provided a word RAM algorithm that outputs all runs in linear time \cite{Kolpakov1999}. 
The algorithm is based on the Lempel-Ziv factorization and only achieves $\orderof{n}$ time for \emph{linearly-sortable alphabets}, i.e.\ alphabets that are totally ordered and for which a sequence of $\sigma$ alphabet symbols can be sorted in $\orderof{\sigma}$ time. 
Since then, it has been an open question whether there exists a linear time runs algorithm for \emph{general ordered alphabets}, i.e.\ totally ordered alphabets for which the order of any two symbols can be determined in constant time.
Any such algorithm must not use the Lempel-Ziv factorization, since for general ordered alphabets of size $\sigma$ it cannot be constructed in $o(n \lg \sigma)$ time \cite{Kosolobov2015}.

Kolpakov and Kucherov also conjectured that the maximum number of runs is bounded by $\rho(n) < n$, which started a 15~year-long search for tighter upper bounds of $\rho(n)$.
Rytter was the first to give an explicit constant with $\rho(n) < 5n$ \cite{Rytter2006}. 
After multiple incremental improvements of this bound (e.g.\ \cite{Puglisi2008,Crochemore2008,Crochemore2011}), Bannai et al.\ \cite{Bannai2017} finally proved the conjecture by showing $\rho(n)<n$ for arbitrary alphabets, which was subsequently even surpassed for binary texts \cite{Fischer2015}.
(The current best bound for binary alphabets is $\rho(n)<\frac{183}{193}n$ \cite{Holub2017}.)

On the algorithmic side, Bannai et al.\ also provided a new linear time algorithm that computes all the runs \cite{Bannai2017}.
While (just like the algorithm by Kolpakov and Kucherov) it only achieves the time bound for linearly-sortable alphabets, it no longer relies on the Lempel-Ziv factorization.
Instead, the main effort of the algorithm lies in the computation of $\Theta(n)$ \emph{longest common extensions (LCEs)}; given two indices $i,j \in [1, n]$, their LCE is the length of the longest common prefix of the suffixes $\str[i..n]$ and $\str[j..n]$.
For linearly-sortable alphabets, we can precompute a data structure in $\orderof{n}$ time that answers arbitrary LCE queries in constant time (see e.g.\ \cite{Fischer2006}), thus yielding a linear time runs algorithm.
Kosolobov showed that for general ordered alphabets any batch of $\orderof{n}$ LCEs can be computed in $\orderof{n \lg^{2/3} n}$ time, and conjectured the existence of a linear time runs algorithm for general ordered alphabets \cite{Kosolobov2016}. 
Gawrychowski et al.\ improved this result to $\orderof{n \lg\lg n}$ time \cite{Gawrychowski2016}. Finally, Crochemore et al.\ noted that the required LCEs satisfy a special non-crossing property. 
They showed how to compute $\orderof{n}$ non-crossing LCEs in $\orderof{n\alpha(n)}$ time, resulting in an $\orderof{n\alpha(n)}$ time algorithm that computes all runs over general ordered alphabets \cite{Crochemore2016} (where $\alpha$ is the inverse Ackermann function).
\subparagraph*{Our Contributions.} We show how to compute the required LCEs in $\orderof{n}$ time and space, resulting in the first linear time runs algorithm for general ordered alphabets, and thus proving Kosolobov's conjecture.
Our solution differs from all previous approaches in the sense that it cannot answer a sequence of \emph{arbitrary} non-crossing LCE queries.
Instead, our algorithm is specifically designed \emph{exactly for the LCEs required by the runs algorithm}.
This allows us to utilize powerful combinatorial properties of the \emph{Lyndon array} (a definition follows in \cref{sec:prelim}) that do not generally hold for arbitrary non-crossing LCE sequences.

Even though the main contribution of our work is the improved asymptotic time bound, it is worth mentioning that our algorithm is also very fast in practice. On modern hardware, computing all runs for a text of length $10^7 (=10\text{MB})$ takes only one second.

\subparagraph*{A Note on the Model.}
As mentioned earlier, our algorithm runs in linear time for \emph{general ordered alphabets}, whereas previous algorithms achieve this time bound only when the alphabet is \textit{linearly-sortable}.
This is comparable with the distinction between \textit{comparison-based} and \textit{integer} sorting: while in the comparison-model sorting $n$ items requires $\Omega(n\lg n)$ time, integer sorting is faster ($O(n\sqrt{\lg\lg n})$ time \cite{Han2002} and sometimes even linear, e.g.\ when the word width $w$ satisfies $w = \orderof{\lg n}$ and one can use radix sort, or when $w \geq (\lg^{2 + \epsilon} n)$ \cite{Andersson1998}).
Whereas it is a major open problem whether integer sorting can always be done in linear time, this paper settles a symmetric open problem for the computation of runs. 

\vspace{.5\baselineskip}

The remainder of the paper is structured as follows: First, we introduce the basic notation, definitions, and auxiliary lemmas (\cref{sec:prelim}). 
Then, we give a simplified description of the runs algorithm by Bannai et al.\ and show how the required LCEs relate to the Lyndon array (\cref{sec:bannairevisit}).
Our linear time algorithm to compute the LCEs is described in \cref{sec:lcealgo}.
We discuss additional practical aspects and experimental results in \cref{sec:practical}.

\section{Preliminaries}
\label{sec:prelim}

Our analysis is performed in the word RAM model (see e.g.\ \cite{Hagerup1998}), where we can perform fundamental operations (logical shifts, basic arithmetic operations etc.) on words of size $w$ bits in constant time.
For an input of size $n$ we assume $\ceil{\log_2 n} \leq w$. We write $[i,j] = [i, j + 1) = (i - 1, j] = (i-1, j+1)$ with $i,j \in \mathbb{N}$ to denote the set of integers ${\{x \mid x \in \mathbb{N} \land i \leq x \leq j \}}$. 

\subparagraph*{Strings.} Let $\Sigma$ be a finite and totally ordered set.
A \emph{string} $\str$ of length $\absolute{\str} = n$ over the \emph{alphabet} $\Sigma$ is a sequence $\str[1]\dots\str[n]$ of $n$ \emph{symbols} (also called \emph{characters}) from $\Sigma$. The alphabet is called a \emph{general ordered alphabet} if order testing (i.e.\ evaluating $\sigma_1 < \sigma_2$ for $\sigma_1, \sigma_2 \in \Sigma$) is possible in constant time.
For $i, j \in [1, n]$, we use the interval notation ${\str[i..j]} = {\str[i..j+1)} = {\str(i - 1..j]} = {\str(i - 1..j+1)}$ to denote the \emph{substring} $\str[i]\dots\str[j]$. 
If however $i > j$, then $\str[i..j]$ denotes the \emph{empty string} $\epsilon$. 
The substring $\str[i..j]$ is called \emph{proper} if $\str[i..j] \neq \str$. 
A \emph{prefix} of $\str$ is a substring $\str[1..j]$ (including $\str[1..0] = \epsilon$), while the \emph{suffix} $\str_i$ is the substring $\str[i..n]$ (including $\str_{n + 1} = \epsilon$). 
Given two strings $\str$ and $\stralt$ of length $n$ and $m$ respectively, their concatenation is defined as $\str\stralt = \str[1]\dots\str[n]\stralt[1]\dots\stralt[m]$. For any positive integer $k$, the $k$-times concatenation of $\str$ is denoted by $\str^k$.
Let $\ellmax = \min(n, m)$. 
The \emph{longest common prefix (LCP)} of $S$ and $T$ has length $\lcp{\str, \stralt} = {\max\{ \ell \mid \ell \in [0, \ellmax] \land \str[1..\ell] = \stralt[1..\ell]\}}$, while the \emph{longest common suffix (LCS)} has length $\lcs{\str, \stralt} = \max\{ \ell \mid {\ell \in [0, \ellmax]} \land {\str_{n - \ell + 1} = \stralt_{m - \ell + 1}}\}$. 
Let $\ell' = \lcp{\str, \stralt}$.
For a string $\str$ of length $n$ and indices $i, j \in [1, n]$, we define the \emph{longest common right-extension (R-LCE) and left-extension (L-LCE)} as $\rlce{i, j} = \lcp{\str_i, \str_j}$ and $\llce{i, j} = \lcs{\str[1..i], \str[1..j]}$ respectively%
.
The total order on $\Sigma$ induces a \emph{lexicographical order} $\llex$ on the strings over $\Sigma$ in the usual way. 
Given three suffixes, we can deduce properties of their R\=/LCEs from their lexicographical order:

\begin{lemma}\label{lemma:lex-deduce-lce}
  Let $\str_i \llex \str_j \llex \str_k$ be suffixes a string, then it holds $\rlce{i, k} \leq \rlce{i,j}$ and $\rlce{i,k} \leq \rlce{j, k}$.
  \begin{proof}
  	Assume $\ell = \rlce{i,j} < \rlce{i, k}$, then $\str_i[1..\ell] = \str_j[1..\ell] = \str_k[1..\ell]$ and $\str_j[\ell + 1] \neq \str_i[\ell + 1] = \str_k[\ell + 1]$. This implies $\str_i \llex \str_j \Leftrightarrow \str_k \llex \str_j$, which contradicts $\str_i \llex \str_j \llex \str_k$. The proof of $\rlce{i,k} \leq \rlce{j, k}$ works analogously.
  \end{proof}
\end{lemma}

\subparagraph*{Repetitions and Runs.}
Let $\str$ be a string and let $\str[i..j]$ be a non-empty substring.
We say that $p \in \mathbb{N}^{+}$ is a \emph{period} of $\str[i..j]$ if and only if ${\forall x \in [i, j - p]} : {\str[x] = \str[x + p]}$. 
If additionally $(j - i + 1) \geq p$, then $\str[i..i+p)$ is called \emph{string period} of $\str[i..j]$. 
Furthermore, $p$ is called \emph{shortest period} of $\str[i..j]$ if there is no $q \in [1, p)$ that is also a period of $\str[i..j]$. 
Analogously, a string period of $\str[i..j]$ is called \emph{shortest string period} if there is no shorter string period of $\str[i..j]$. 
A \emph{run} is a triple $\triple{i, j, p}$ such that $p$ is the shortest period of $\str[i..j]$, $(j - i + 1) \geq 2p$ (i.e.\ there are at least two consecutive occurrences of the shortest string period $\str[i..i + p)$), and 
neither $\triple{i - 1, j, p}$ nor $\triple{i, j + 1, p}$ satisfies these properties (assuming $i > 1$ and $j < n$, respectively).
\subparagraph*{Lyndon Words and Nearest Smaller Suffixes.}
For a length-$n$ string $\str$ and $i \in [1, n]$, the string $\str_i\str[1..i)$ is called \emph{cyclic shift} of $\str$, and \emph{non-trivial cyclic shift} if $i > 1$. A \emph{Lyndon word} is a non-empty string that is lexicographically smaller than any of its non-trivial cyclic shifts, i.e.\ $\forall i \in [2,n] : \str \llex \str_i\str[1..i)$.
The Lyndon array of $\str$ identifies the longest Lyndon word starting at each position of $\str$.

\begin{definition}[Lyndon Array]
  Given a string $\str$ of length $n$, its Lyndon array $\lambda[1,n]$ is defined by $\forall i \in [1, n] :\lyndarr{i} = \max\{ j - i + 1 \mid j \in [i,n] \land \str[i..j] \text{ is a Lyndon word} \}$.
\end{definition}

An alternative representation of the Lyndon array is the next-smaller-suffix array.

\begin{definition}[Next Smaller Suffixes]\label{def:xss}
  Given a string $\str$ of length $n$, its \emph{next-smaller-suffix (NSS) array} is defined by ${\forall i \in [1, n]}: {\nss{i} = \min\{j \mid j = n + 1} \lor {(j \in (i, n] \land \str_i \glex \str_j)\}}$.
  If $\nss{i} \leq n$, then $\str_{\nss{i}}$ is called \emph{next smaller suffix of $\str_i$}.
\end{definition}

\begin{lemma}[Lemma 15 \cite{Franek2016}]\label{lemma:lyndonnss}
  The longest Lyndon word starting at any position $i \in [1, n]$ of a length-$n$ string $\str$ is exactly the substring $\str[i..\nss{i})$, i.e.\ $\forall i \in [1, n] : {\lyndarr{i} = \nss{i} - i}$.
\end{lemma}

An important property of next smaller suffixes is that they do not \emph{intersect}:

\begin{lemma}\label{lemma:nonintersecting}
	Let $i \in [1, n]$ and $i' \in [i, \nss{i})$. Then it holds $\nss{i'} \leq \nss{i}$. 
	\begin{proof}
		Due to $i' \in [i, \nss{i})$ and \cref{def:xss} it holds $\str_{i'} \glex \str_{\nss{i}}$.
		Assume that the lemma does not hold, then we have $\nss{i} \in (i', \nss{i'})$ and \cref{def:xss} implies $\str_{i'} \llex \str_{\nss{i}}$.
	\end{proof}
\end{lemma}

\section{The Runs Algorithm Revisited}
\label{sec:bannairevisit}

In this section, we recapitulate the main ideas of the algorithm by Bannai et al. \cite{Bannai2017}, which is the basis of our solution for general ordered alphabets.
The key insight is that every run is \emph{rooted} in a longest Lyndon word, allowing us to compute all runs from the Lyndon array.

\begin{definition}\label{def:incdec}
  Let $\triple{i, j, p}$ be a run in a string $\str$. 
  We say that $\triple{i, j, p}$ is \emph{(lexicographically) decreasing} if and only if $\str_i \glex \str_{i + p}$. 
  Otherwise, $\triple{i, j, p}$ is \emph{(lexicographically) increasing}.
\end{definition}

\begin{lemma}\label{lemma:lyndonroots}
  Let $\triple{i, j, p}$ be a decreasing run, then there is exactly one index ${i_0 \in [i..i+p)}$ such that $\lyndarr{i_0} = p$.
  \begin{proof}
    Consider any $i_0 \in [i, i + p)$. By the definition of runs, we have $\str[i..i_0) = \str[i + p..i_0 + p)$. Since the run is decreasing it follows 
    $
      {\str_i \glex \str_{i + p}} 
      \iff {\str[i..i_0)\str_{i_0} \glex \str[i + p..i_0+p)\str_{i_0 + p}}
      \iff {\str_{i_0} \glex \str_{i_0 + p}}
    $. 
    This implies $\nss{i_0} \leq i_0 + p$, and due to \cref{lemma:lyndonnss} also $\lyndarr{i_0} \leq p$.   
    Next, we show that there is at least one index $i_0 \in [i..i+p)$ such that $\str[i_0..i_0 + p)$ is a Lyndon word.
    Let $\alpha = \str[i..i+p)$. 
    Assume that the described index $i_0$ does not exist, then from $\str[i..i + 2p) = \alpha\alpha$ follows that no cyclic shift of $\alpha$ is a Lyndon word.
    Let $\beta$ be a lexicographically minimal cyclic shift of $\alpha$, then this shift is not unique (otherwise $\beta$ would be a Lyndon word), and thus there must be a cyclic shift $\beta_k\beta[1..k) = \beta[1..k)\beta_k$ with $k > 1$.
    This however implies that $\beta$ is of the form $\beta = \mu^k$ for some string $\mu$ and an integer $k > 1$ (see Lemma 3 in \cite{Lyndon1962}), which contradicts the fact that $\alpha$ is the \emph{shortest} string period of the run. 
    Finally, let $\alpha_{k}\alpha[1..k)$ with $k \in [1,p]$ be the unique lexicographically smallest cyclic shift of $\alpha$ (and thus a Lyndon word), then it is easy to see that only $i_0 = i + k - 1$ satisfies $\lyndarr{i_0} = p$.
  \end{proof}
\end{lemma}

\begin{definition}[Root of a Run]
  Let $\triple{i, j, p}$ be a decreasing run, and let ${i_0 \in [i..i+p)}$ be the unique index with $\lyndarr{i_0} = p$ (as described in \cref{lemma:lyndonroots}). We say that $\triple{i, j, p}$ is \emph{rooted in} $i_0$.
\end{definition}

\begin{figure}[t]
\centering
\newcommand{\tmp}{\vphantom{[]}}
\newcommand{\s}{\hspace{2\strboxspace}\stringstrut}
\newcommand{\tmps}{\hspace{2\strboxspace}}

\rlap{\color{white}.}\parbox[t]{.72\textwidth}{
\vspace{1.5\baselineskip}
$%
\begin{aligned}
\str =\ &\stringbox[0pt]{}{\examplestr{aaaa}\s%
{\overbrace{%
{\overbrace{\tikzmark{ul1}\examplestr{abc}\tikzmark{ul2}{\s}\tikzmark{n1}\tikzmark{ul7}\examplestr{abab}}^{\alpha\tmp}}\s
{\overbrace{\tikzmark{ul3}\examplestr{abc}\tikzmark{ul4}{\s}\tikzmark{n2}\tikzmark{ul5}\examplestr{abab}}^{\alpha\tmp}}\s
{\overbrace{\examplestr{abc\s}\tikzmark{n3}\examplestr{aba}\tikzmark{ul8}\examplestr{b}}^{\alpha\tmp}}\s
{\overbrace{\examplestr{abc\s}\tikzmark{n4}\examplestr{aba}\tikzmark{ul6}}^{\alpha[1..6]}}}^{\displaystyle\text{run }\triple{5,31,7}}}\s
\examplestr{aaaa}}{}\\[.91\baselineskip]
&\phantom{\stringbox[0pt]{}{\examplestr{aaaa\s abc}}{}}%
\mathllap{\scriptstyle\beta\ =\ }%
\stringbox[0pt]{}{\tikzmark{n1}\examplestr{abab\s abc}}{}
\hspace{\strboxspace}\tikzmark{n2}
\end{aligned}
\begin{tikzpicture}[overlay, remember picture]
  \path (n1) ++(2pt,-3pt) node (n1) {};
  \path (n2) ++(2pt,-3pt) node (n2) {};
  \draw[black, thick, -latex] (n1.center) to[out=270, in=270] node[pos=.9, right, align=center] {\ $\nss{8} = 15,\ \lyndarr{8} = 7$} (n2.center) to ++(0pt, 7pt);
  \foreach \x in {1,3,5,7} {
	  \path (ul\x.center) ++(-.3mm, -2.1mm) node (ul\x) {};
  }
  \foreach \x in {2,4,6,8} {
	  \path (ul\x.center) ++(.3mm, -2.1mm) node (ul\x) {};
  }
  \foreach \x in {2,4} {
	  \path (ul\x.center) ++(2.5mm, 0) node (ul\x) {};
  }
  \foreach \x in {5,6,7,8} {
	  \path (ul\x.center) ++(0, -5pt) node (ul\x) {};
  }
  \draw[my-second, line width=3.5pt] (ul1.center) -- (ul2.center);
  \draw[my-second, line width=3.5pt] (ul3.center) -- (ul4.center);
  \draw[my-first, line width=3.5pt] (ul5.center) -- (ul6.center);
  \draw[my-first, line width=3.5pt] (ul7.center) ++(0, -5pt) -- ($(ul8.center)+(0, -5pt)$);
\end{tikzpicture}
$
}\hfill\parbox[t]{.27\textwidth}{\textcolor{white}{.}%

\vspace{.5\baselineskip}

\tikzmark{label1}$\llce{8, 15} = 4$

\tikzmark{label2}$\rlce{8, 15} = 17$

\begin{tikzpicture}[overlay, remember picture]
  \draw[my-second, line width=3.5pt] (label1.center) ++(-1mm, .8mm) -- ++(-7mm, 0);
  \draw[my-first, line width=3.5pt] (label2.center) ++(-1mm, .8mm) -- ++(-7mm, 0);
\end{tikzpicture}
}

\vspace{.75\baselineskip}

\caption{Lexicographically decreasing run $\triple{5,31,7}$ with $\str[5..31] = (\texttt{abcabab})^{27/7}$. The run has shortest string period $\alpha = \examplestr{abcabab}$, and is rooted in position $8$ (with longest Lyndon word $\beta = \str[8..15) = \alpha_4\alpha[1..3] = \examplestr{abababc}$).}
\label{fig:runexample}
\end{figure}
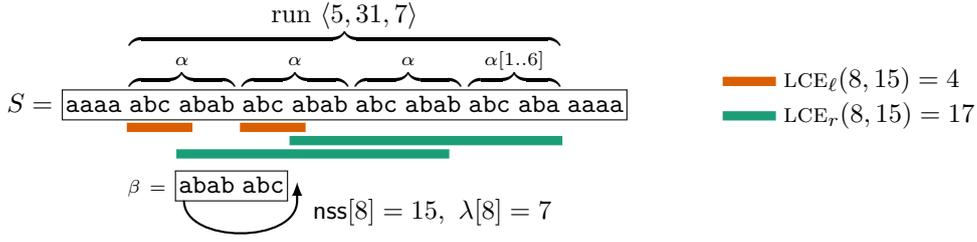

An example of a decreasing run and its root is provided in \cref{fig:runexample}. 
Note that our notion of a root differs from the \textsf{L}-roots introduced by Crochemore et al.\ \cite{Crochemore2014}. While an \textsf{L}-root is \emph{any} length-$p$ Lyndon word contained in the run, our root is exactly the \emph{leftmost} one.

Given a longest Lyndon word $S[i_0..\nss{i_0})$ of length $p = \nss{i_0} - i_0 = \lyndarr{i_0}$, it is easy to determine whether $i_0$ is the root of a decreasing run.
We simply try to extend the periodicity as far as possible to both sides by using the LCE functions. 
For this purpose, we only need to compute $l = \llce{i_0, \nss{i_0}}$ and $r = \rlce{i_0, \nss{i_0}}$.
Let $i = i_0 - l + 1$ and $j = \nss{i_0} + r - 1$, then clearly the substring $\str[i..j]$ has smallest period $p$, and we cannot extend the substring to either side without breaking the periodicity.
Thus, if $j - i + 1 \geq 2p$ then $\triple{i, j, p}$ is a run.
Note that this run is only rooted in $i_0$ if additionally $i_0 \in [i..i + p)$ (or equivalently $l \leq p$) holds.
For the index $i_0 = 8$ in \cref{fig:runexample}, we have $l = \llce{8, 15} = 4$ and $r = \rlce{8, 15} = 17$. 
Therefore, the run starts at position $i = 8 - 4 + 1 = 5$ and ends at position $j = 15 + 17 - 1 = 31$. 
From $l = 4 \leq 7 = p$ follows that $8$ is actually the root.

Since each decreasing run is rooted in exactly one index, we can find all decreasing runs by checking for each index whether it is the root of a run.
This procedure is outlined in \cref{alg:highlevel}.
First, we compute the NSS array (line \ref{alg:highlevel:computenss}).
Then, we investigate one index $i_0 \in [1, n]$ at a time (line \ref{alg:highlevel:loop}), and consider it as the root of a run with period $p = \nss{i_0} - i_0$ (line \ref{alg:highlevel:period}).
If the left-extension covers an entire period (i.e.\ $\llce{i_0, \nss{i_0}} > p$), then we have already investigated the root of the run in an earlier iteration of the for-loop, and no further action is required (line \ref{alg:highlevel:checkllce}).
Otherwise, we compute the left and right border of the potential run as described earlier (lines \ref{alg:highlevel:llce}--\ref{alg:highlevel:rlce}). 
If the resulting interval has length at least $2p$, then we have discovered a run that is rooted in $i_0$ (lines \ref{alg:highlevel:checktwoperiods}--\ref{alg:highlevel:addrun}).

\begin{algorithm}
\begin{algorithmic}[1]
  \Require{String $\str$ of length $n$.}
  \Ensure{Set $R$ of all decreasing runs in $\str$.}
  \State $R \gets \emptyset$
  \State compute array \textsf{nss}\label{alg:highlevel:computenss}
  \For{$i_0 \in [1, n]$\textbf{ with }$\nss{i_0} \neq n + 1$}\label{alg:highlevel:loop}
  \State $p \gets \nss{i_0} - i_0$ \label{alg:highlevel:period}
  \If{$\llce{i_0, \nss{i_0}} \leq p$} \label{alg:highlevel:checkllce}
  \State $\phantom{j}\mathllap{i} \gets \phantom{\nss{i_0}}\mathllap{i_0\phantom{]}} - \llce{i_0, \nss{i_0}} + 1$ \label{alg:highlevel:llce}
  \State $j \gets \nss{i_0} + \rlce{i_0, \nss{i_0}} - 1$\label{alg:highlevel:rlce}
  \If{$j - i + 1 \geq 2p$}\label{alg:highlevel:checktwoperiods}
  \State $R \gets R \cup \{ \triple{i, j, p} \}$\label{alg:highlevel:addrun}
  \EndIf
  \EndIf
  \EndFor
\end{algorithmic}
\caption{Compute all decreasing runs.}
\label{alg:highlevel}
\end{algorithm}

\subparagraph*{Time and space complexity.} 
The NSS array can be computed in $\orderof{n}$ time and space for general ordered alphabets \cite{Bille2020}. 
Assume for now that we can answer L-LCE and R-LCE queries in constant time, then clearly the rest of the algorithm also requires $\orderof{n}$ time and space. The correctness of the algorithm follows from \cref{lemma:lyndonroots} and the description. We have shown:

\begin{lemma}\label{lemma:requiredlces}
	Let $\str$ be a string of length $n$ over a general ordered alphabet, and let $\textnormal{\textsf{nss}}$ be its NSS array. We can compute all increasing runs of $\str$ in $\orderof{n} + t(n)$ time and $\orderof{n} + s(n)$ space, where $t(n)$ and $s(n)$ are the time and space needed to compute $\llce{i, \nss{i}}$ and $\rlce{i, \nss{i}}$ for all $i \in [1, n]$ with $\nss{i}\neq n + 1$.
\end{lemma}

In order to also find all \emph{increasing} runs, we only need to rerun the algorithm with reversed alphabet order.
This way, previously increasing runs become decreasing.

\section{Algorithm for Computing the LCEs}
\label{sec:lcealgo}

In this section, we show how to precompute the LCEs required by \cref{alg:highlevel} in linear time and space. Our approach is asymmetric in the sense that we require different algorithms for L\=/LCEs and R\=/LCEs (whereas previous approaches usually compute L\=/LCEs by applying the R-LCE algorithm to the reverse text). However, for both directions we use similar properties of the Lyndon array that are shown in \cref{lemma:lyndonchainright,lemma:lyndonchainleft} and visualized in \cref{fig:lyndonchain}.

\begin{lemma}\label{lemma:lyndonchainright}
	Let $i \in [1, n]$ and $j = \nss{i} \neq n + 1$. If $\rlce{i, j} \geq (j - i)$, then it holds  $\rlce{j, j + (j - i)} = \rlce{i, j} - (j - i)$ and $\nss{j} = j + (j - i)$.
	\begin{proof}
		From $\rlce{i, j} \geq (j - i)$ follows $\rlce{i, j} = (j - i) + \rlce{j, j + (j - i)}$, which is equivalent to $\rlce{j, j + (j - i)} = \rlce{i, j} - (j - i)$.
		It remains to be shown that $\nss{j} = j + (j - i)$.
		Due to $\nss{i} = j$ it holds $\str_i \glex \str_j$.
		Since ${\str_i \glex \str_j}$ and $\rlce{i, j} \geq (j - i)$, we have $\str_{i + (j - i)} \glex \str_{j + (j - i)}$, which implies $\nss{j} \leq j + (j - i)$.
		Note that $\nss{i} = j$ and \cref{lemma:lyndonnss} imply that $\str[i..j) = \str[j..j+(j - i))$ is a Lyndon word. Thus it holds $\lyndarr{j} \geq (j - i)$, or equivalently $\nss{j} \geq j + (j - i)$.
	\end{proof}
\end{lemma}

\begin{lemma}\label{lemma:lyndonchainleft}
	Let $i \in [1, n]$ and $j = \nss{i} \neq n + 1$. If $\llce{i, j} > (j - i)$, then it holds $\llce{i - (j - i), i} = \llce{i, j} - (j - i)$ and $\nss{i - (j - i)} = i$.
	\begin{proof}
		Analogous to \cref{lemma:lyndonchainright}.
	\end{proof}
\end{lemma}

\subsection{Computing the R\=/LCEs}
\label{sec:rlce}

First, we will briefly describe our general technique for computing LCEs, and our method of showing the linear time bound. 
Assume for this purpose that we want to compute $\ell = \rlce{i,j}$ with $i < j$. 
It is easy to see that we can determine $\ell$ by performing $\ell + 1$ individual character comparisons (by simultaneously scanning the suffixes $\str_i$ and $\str_j$ from left to right until we find a mismatch).
Whenever we use this naive way of computing an LCE, we \emph{charge} one character comparison to each of the indices from the interval $[j, j + \ell)$. 
This way, we account for $\ell$ character comparisons. 
Since we want to compute $\orderof{n}$ R-LCE values in $\orderof{n}$ time, we can afford a constant time overhead (i.e.\ a constant number of unaccounted character comparisons) for each LCE computation.
Thus, there is no need to charge the $(\ell + 1)$-th comparison to any index.
At the time at which we want to compute~$\ell$, we may already know some lower bound $k \leq \ell$. 
In such cases, we simply skip the first $k$ character comparisons and compute $\ell = k + \rlce{i + k, j + k}$. 
This requires $\ell - k + 1$ character comparisons, of which we charge $\ell - k$ to the interval $[j + k..j+\ell)$.

\begin{figure}[t]
\subcaptionbox{\cref{lemma:lyndonchainright,lemma:lyndonchainleft}. The dotted edge follows from $\rlce{i, j} \geq (j - i)$ (\cref{lemma:lyndonchainright}). The dashed edge follows from $\llce{i, j} > (j - i)$ (\cref{lemma:lyndonchainleft}).\label{fig:lyndonchain}}%
{\parbox[t]{.485\textwidth}{\centering%
\phantom{$I^{I^{\strut I}}$}\\
$\str = \stringbox{}{}{}%
\tikzmark{1}%
\stringbox[.5cm]{\stringidx{i - (j - i)}}{\beta}{}
\tikzmark{e1}\tikzmark{2}%
\stringbox[.5cm]{\stringidx{i}}{\beta}{}
\tikzmark{e2}\tikzmark{3}%
\stringbox[.5cm]{\stringidx{j}}{\beta}{}
\tikzmark{e3}%
\stringbox{\stringidx{j + (j - i)}}{\phantom{a}}{}%
\begin{tikzpicture}[overlay, remember picture]
	\path (1) ++(2pt,-4pt) node (1) {} ++(0,-11pt) node (i1) {};
	\foreach \x in {2,3} {
		\path (\x) ++(3.5pt,-10pt) node (\x) {} ++(0,-5pt) node (i\x) {};
	}
	\foreach \x in {1,2,3} {
		\path (e\x) ++(2pt,-4pt) node (e\x) {} ++(0,-11pt) node (ie\x) {};
	}
	\draw[-Latex, dashed] (1.center) to (i1.center) to[out = 270, in = 270, looseness = .7] (ie1.center) to (e1.center);
	\draw[-Latex] (2.center) to (i2.center) to[out = 270, in = 270, looseness = .7] (ie2.center) to (e2.center);
	\draw[-Latex, dotted] (3.center) to (i3.center) to[out = 270, in = 270, looseness = .7] (ie3.center) to (e3.center);
\end{tikzpicture}$

\vspace{2.25\baselineskip}
}}\hfill%
\subcaptionbox{Relative order of R\=/LCE computations from first to last: 
$\rlce{i_1, j_1}$,
$\rlce{i_2, j_1}$,
$\rlce{i_3, j_2}$,
$\rlce{i_4, j_2}$,
$\rlce{i_5, j_2}$,
$\rlce{i_6, j_2}$.\label{fig:orderofcomputation}}%
{\parbox[t]{.475\textwidth}{\centering%
\phantom{$I^{I^I}$}\\
$\str = \stringbox{}{%
\phantom{a}%
\tikzmark{i6}%
\stringidx{i_6}\phantom{aaa}%
\tikzmark{i5}%
\stringidx{i_5}\phantom{aa}%
\tikzmark{i2}%
\stringidx{i_2}\phantom{aaaa}%
\tikzmark{i1}%
\stringidx{i_1}\phantom{aa}%
\tikzmark{j1}\tikzmark{j2}\tikzmark{i4}%
\stringidx{j_1}\phantom{aa}%
\tikzmark{i4}%
\stringidx{i_4}\phantom{aa}%
\tikzmark{i3}%
\stringidx{i_3}\phantom{aaa}%
\tikzmark{j3}\tikzmark{j4}\tikzmark{j5}\tikzmark{j6}%
\stringidx{j_2}}{}
\begin{tikzpicture}[overlay, remember picture]
	\foreach \x in {1,...,6} {
		\path (i\x) ++(0,-4pt) node (i\x) {} ++(0,-6pt) node (ii\x) {};
		\path (j\x) ++(0,-4pt) node (j\x) {} ++(0,-6pt) node (jj\x) {};
	}
	\foreach \x in {1,...,6} {
		\draw[-Latex] (i\x.center) to (ii\x.center) to[out = 270, in = 270, looseness = .4] node (lab\x) {} (jj\x.center) to (j\x.center);
	}
\end{tikzpicture}$

\vspace{2.25\baselineskip}}}
\caption{An edge from text position $a$ to text position $b$ indicates $\nss{a} = b$.}
\end{figure}

\vspace{.5\baselineskip}

Ultimately, we will show that all R\=/LCE values $\rlce{i, j}$ with $i \in [1, n]$ and $j = \nss{i} \neq n + 1$ can be computed in a way such that each text position gets charged at most once, which results in the desired linear time bound. 
From now on, we refer to $i$ as the \emph{left index} and $j$ as the \emph{right index} of the R\=/LCE computation.
Our algorithm computes the R\=/LCEs in the following order (a visualization is provided in \cref{fig:orderofcomputation}): 
We consider the possible right indices $j \in [2, n]$ one at a time and in \emph{increasing} order. 
For each right index $j$, we then consider the corresponding left indices $i$ with $\nss{i} = j$ in \emph{decreasing} order (we will see how to efficiently deduce this order from the Lyndon array later).

Assume that we are computing the R\=/LCEs in the previously described order, and let $\ell = \rlce{i, j}$ with $j = \nss{i} \neq n + 1$ be the next value that we want to compute. The set of indices that we have already considered as left indices for LCE computations is $I = \{ x \mid (\nss{x} < j) \lor ((\nss{x} = j) \land (i < x))\}$. For example, when we compute $\rlce{i_4, j_2}$ in \cref{fig:orderofcomputation} it holds $\{i_1, i_2, i_3\} \subseteq I$. At this point in time, the rightmost text position that we have already inspected is $\rhs = \max_{x \in I}(\nss{x} + \rlce{x, \nss{x}})$ if $I \neq \emptyset$, or $\rhs = 1$ otherwise.
%
%
%
Due to the nature of our charging method, we have not charged any indices from the interval $[\rhs, n]$ yet. 
Thus, in order to show that we can compute all LCEs without charging any index twice, it suffices to show how to compute $\ell = \rlce{i, j}$ without charging any index from the interval $[1, \rhs)$.
If $j \geq \rhs$ then we naively compute $\ell$ and charge the character comparisons to the interval $[j, j+\ell)$, thus only charging previously uncharged indices.
The new value of $\rhs$ is $j + \ell$.
If however $j < \rhs$, then the computation of $\ell$ depends on the previously computed LCEs, which we describe in the following.

Let $\ell' = \rlce{i', j'}$ with $j' = \nss{i'}$ be the \emph{most recently} computed R\=/LCE that satisfies $j' + \ell' = \rhs$. 
Our strategy for computing $\ell$ depends on the position of $i$ relative to $i'$ and $j'$. 
First, note that $i \notin [i', j')$ because otherwise
\cref{lemma:nonintersecting} implies $j \leq j'$, which contradicts our order of computation.
This leaves us with three possible cases (as before, a directed edge from text position $a$ to text position $b$ indicates $\nss{a} = b$): 

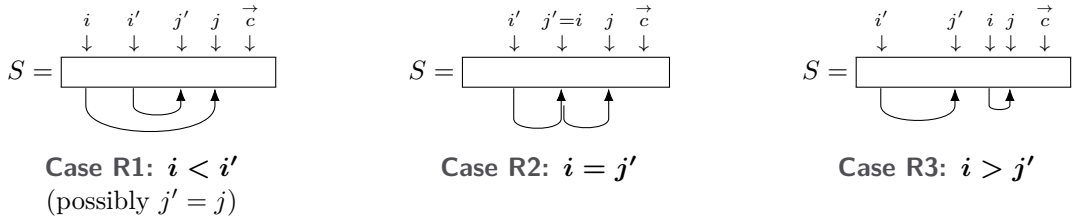
\begin{center}
\noindent\shortstack{\vphantom{\shortstack{{I}\\{I}\\{I}}}$\str=\stringbox{}{%
\tikzmark{s1}\stringidx{i}\quad\enskip%
\tikzmark{s2}\stringidx{i'}\quad\enskip%
\tikzmark{t2}\stringidx{j'}\quad%
\tikzmark{t1}\stringidx{j}\quad%
\stringidx{\rhs}}{}%
\begin{tikzpicture}[overlay, remember picture]
	\foreach \x in {1,2} {
		\path (s\x) ++(1pt, 0) node (s\x) {};
		\path (t\x) ++(1pt, 0) node (t\x) {} ++(0,-10pt) node (m\x) {};
		\draw[-Latex] (s\x) to ++(0, -10pt) to[out=270, in=270, looseness=.7] (m\x.center) to (t\x);
	}
\end{tikzpicture}
$\\${\strut}^{\strut}$\\\shortstack{\textbf{\case{R1} \bm{i < i'}}\vphantom{j}\\(possibly $j' = j$)}}%
\hfill%
\shortstack{$\str=\stringbox{}{%
\quad%
\tikzmark{s2}\stringidx{i'}\quad\enskip%
\tikzmark{t2}\tikzmark{s1}\stringidx{j' = i}\quad\enskip%
\tikzmark{t1}\stringidx{j}\quad%
\stringidx{\rhs}}{}%
\begin{tikzpicture}[overlay, remember picture]
	\foreach \x in {1,2} {
		\path (s\x) ++(1pt, 0) node (s\x) {};
		\path (t\x) ++(1pt, 0) node (t\x) {} ++(0,-15pt) node (m\x) {};
	}
	\draw[-Latex] (s2) to ++(0, -15pt) to[out=270, in=270, looseness=.7] (m2.center) to (t2);
	\draw[-Latex] (s1) ++(1pt, -10pt) to ++(0,-5pt) to[out=270, in=270, looseness=.7] (m1.center) to (t1);
\end{tikzpicture}
$\\${\strut}^{\strut}$\\\shortstack{\textbf{\case{R2} \bm{i = j'}}\\\phantom{(y}}}%
\hfill%
\shortstack{$\str=\stringbox{}{%
\tikzmark{s2}\stringidx{i'}\quad\quad\enskip%
\tikzmark{t2}\stringidx{j'}\quad%
\tikzmark{s1}\stringidx{i}\enskip%
\tikzmark{t1}\stringidx{j}\quad%
\stringidx{\rhs}}{}%
\begin{tikzpicture}[overlay, remember picture]
	\foreach \x in {1,2} {
		\path (s\x) ++(1pt, 0) node (s\x) {};
		\path (t\x) ++(1pt, 0) node (t\x) {} ++(0,-10pt) node (m\x) {};
		\draw[-Latex] (s\x) to ++(0, -10pt) to[out=270, in=270, looseness=.7] (m\x.center) to (t\x);
	}
\end{tikzpicture}
$\\${\strut}^{\strut}$\\\shortstack{\textbf{\case{R3} \bm{i > j'}}\\\phantom{(y}}}%
\end{center}

Now we explain the cases in detail.
Each case is accompanied by a schematic drawing. 
We strongly advise the reader to study the drawings alongside the description, since they are essential for an easy understanding of the matter.

\vspace{.5\baselineskip}

\newcommand{\tmpa}{.6em}
\newcommand{\tmpb}{.3em}
\newcommand{\tmpc}{1.75mm}
\newcommand{\tmpd}{2em}
\newcommand{\alphabox}[1]{\stringbox[\tmpa]{\stringidx{#1}}{\alpha}{}}
\newcommand{\betabox}[1]{\stringbox[\tmpb]{}{\beta}{}}


\newtcolorbox{itembox}[1][]{enhanced jigsaw, breakable=true, left=0mm,top=1mm,right=0mm,bottom=1.5mm, interior hidden, sharp corners=all,
colframe=black!30!white,nobeforeafter=,
#1}


\newenvironment{nitembox}{\parindent0pt{\textcolor{black!30!white}{\hrulefill}\\[.5\baselineskip]}}{}
  
  \begin{itembox}
  \case{R1} \bm{i < i'} (and $j' \leq j < \rhs$)\textbf{.}
  
  \vspace{\baselineskip}
  
  $\absolute{\alpha} = j - j'$, $\absolute{\beta} = \rhs - j$\hfill%
  $\stringstrut%
\str = %
\stringbox[\tmpc]{}{}{}%
\tikzmark{m1}%
\stringbox[\tmpb]{\stringidx{i}}{\beta}{}%
\stringbox[.5mm]{}{\gamma}{}%
\stringbox[\tmpc]{}{}{}%
\tikzmark{m2}%
\alphabox{i'}\stringbox[\tmpb]{\stringidx{\ \ (i' + j - j')}}{\beta}{}%
\stringbox[3mm]{}{}{}%
\tikzmark{m3}%
\alphabox{j'}
\tikzmark{m4}%
\stringbox[\tmpb]{\stringidx{j}}{\beta}{}%
\stringbox[.5mm]{\stringidx{\rhs}}{\gamma}{}%
\stringbox[\tmpc]{}{}{}%
\begin{tikzpicture}[overlay, remember picture]
  \foreach \x in {1,2,3,4} {
    \path (m\x) ++(3pt,-0pt) node (m\x) {}; 
  }

  \draw (m1) edge[looseness = .5, out = 270, in = 270, -Latex] (m4);
  \draw (m2) edge[looseness = .6, out = 270, in = 270, -Latex] (m3);
    
\end{tikzpicture}$

  $\ell' = \absolute{\alpha\beta}$, $\ell = \absolute{\beta\gamma}$
  
  \vspace{1.25\baselineskip}
  
  Due to $i < (i' + j - j') < j = \nss{i}$ we have $\str_j \llex \str_{i} \llex \str_{i' + j - j'}$. 
  From \cref{lemma:lex-deduce-lce} follows $\rhs - j = \rlce{i' + j - j', j} \leq \rlce{i, j} = \ell$, i.e.\ both $\str_i$ and $\str_j$ start with~$\beta$.
  Since now we know a lower bound $\rhs - j \leq \ell$ of the desired LCE value, we can skip character comparisons during its computation. 
  Later, we will see that the same bound also holds for most of the other cases. Generally, whenever we can show $\rhs - j \leq \ell$ we use the following strategy.
  We compute ${\ell = (\rhs - j) + \rlce{i + (\rhs - j), \rhs}}$ using $\ell - (\rhs - j) + 1$ character comparisons, of which we charge $\ell - (\rhs - j)$ to the interval $[\rhs,j+\ell)$. 
  Thus we only charge previously uncharged positions. 
  We continue with $i' \gets i$, $j' \gets j$, $\ell' \gets \ell$, and $\rhs \gets j + \ell$.
  \end{itembox}
  \begin{itembox}
  \case{R2} \bm{i = j'}\textbf{.} We divide this case into two subcases.
  
  {\textcolor{black!30!white}{\hrulefill}\\[-.5\baselineskip]}
  
  \case{R2a} \bm{\ell' < j' - i'}.
  
  \vspace{\baselineskip}
  
  $\absolute{\alpha} = j - j'$, $\absolute{\beta} = \rhs - j$\hfill%
  $\stringstrut%
\str = %
\stringbox[.2cm]{}{}{}%
\tikzmark{m1}%
\stringbox[.35cm]{\stringidx{i'}}{\alpha}{}%
\stringbox[.2cm]{\stringidx{(i' + j - i)}}{\beta}{}%
\stringbox[.3cm]{}{}{}%
\tikzmark{m2}\tikzmark{m3}%
\stringbox[.35cm]{\stringidx{j' = i}}{\alpha}{}%
\tikzmark{m4}%
\stringbox[.2cm]{\stringidx{j}}{\beta}{}%
\stringbox[.2cm]{\stringidx{\rhs}}{}{}%
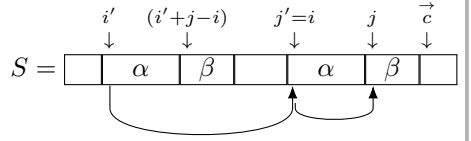
\begin{tikzpicture}[overlay, remember picture]
  \path (m2) ++(-1pt, 0) node (m2) {};
  \foreach \x in {1,2,3,4} {
    \path (m\x) ++(3pt,-0pt) node (m\x) {} ++(0, -12pt) node (im\x) {}; 
  }
  \path (m3) ++(0, -6.5pt) node (m3) {};

  \draw[-Latex] (m1) to (im1.center) to[looseness = .5, out = 270, in = 270, -Latex] (im2.center) to (m2);
  \draw[-Latex] (m3) to (im3.center) to[looseness = .5, out = 270, in = 270, -Latex] (im4.center) to (m4);
    
\end{tikzpicture}$

  \phantom{$\ell = \ell' + \absolute{\beta}$}
  
  \vspace{1.25\baselineskip}
  
  From $j < \rhs \implies j - i < \rhs - i = \ell'$ and $\ell' < j' - i'$ follows $i' + j - i < j' = i$. 
  Therefore, $\nss{i'} = i$ and \cref{def:xss} imply $\str_{i} \llex \str_{i' + j - 1}$. 
  Due to $\nss{i} = j$ we also have $\str_j \llex \str_{i}$, such that it holds $\str_j \llex \str_{i} \llex \str_{i' + j - 1}$. 
  It is easy to see that $\str_{i' + j - i}$ and $\str_j$ share a prefix $\beta$ of length $\rlce{i' + j - i, j} = \rhs - j$.
  In fact, also $\str_i$ has prefix $\beta$ because \cref{lemma:lex-deduce-lce} implies that $\rlce{i' + j - i, j} \leq \rlce{i, j} = \ell$. 
  Thus it holds $\rhs - j \leq \ell$, which allows us to use the strategy from Case R1.
  
  {\textcolor{black!30!white}{\hrulefill}\\[-.5\baselineskip]}

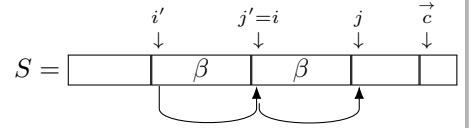
  \case{R2b} \bm{\ell' \geq j' - i'}.
  
  \vspace{\baselineskip}
  
  $\absolute{\beta} = j' - i'$, $\ell = \ell' - \absolute{\beta}$\hfill%
  $\stringstrut%
\str = %
\stringbox[.5cm]{}{}{}%
\tikzmark{m1}%
\stringbox[.5cm]{\stringidx{i'}}{\beta}{}%
\tikzmark{m2}\tikzmark{m3}%
\stringbox[.5cm]{\stringidx{j' = i}}{\beta}{}%
\tikzmark{m4}%
\stringbox[.4cm]{\stringidx{j}}{}{}%
\stringbox[.2cm]{\stringidx{\rhs}}{}{}%
\begin{tikzpicture}[overlay, remember picture]
  \path (m2) ++(-1pt, 0) node (m2) {};
  \foreach \x in {1,2,3,4} {
    \path (m\x) ++(3pt,-0pt) node (m\x) {} ++(0, -12pt) node (im\x) {}; 
  }
  \path (m3) ++(0, -6.5pt) node (m3) {};

  \draw[-Latex] (m1) to (im1.center) to[looseness = .5, out = 270, in = 270, -Latex] (im2.center) to (m2);
  \draw[-Latex] (m3) to (im3.center) to[looseness = .5, out = 270, in = 270, -Latex] (im4.center) to (m4);
    
\end{tikzpicture}$

  \phantom{$\ell = \ell' + \absolute{\beta}$}
  
  \vspace{1.25\baselineskip}
  
  Due to $\ell' \geq j' - i'$, \cref{lemma:lyndonchainright} implies $j = i + (j' - i')$ and $\ell = \ell' - (j' - i')$. Since $i'$, $j'$, and $\ell'$ are known, we can compute $\ell$ in constant time without performing any character comparisons. We continue with $i' \gets i$, $j' \gets j$, and $\ell' \gets \ell$ (leaving $\rhs$ unchanged).
  
  \end{itembox}
  \begin{itembox}%
  \case{R3} \bm{i > j'}.
  This is the most complicated case, and it is best explained by dividing it into three subcases.
  Let $d = j' - i'$, $i'' = i - d$, $j'' = j - d$, and $\ell'' = \rlce{i'', j''}$.
  
  {\textcolor{black!30!white}{\hrulefill}\\[-.5\baselineskip]}   
    
  \case{R3a} \textbf{\boldmath$\nss{i''} \neq j''$\unboldmath:}%
  
  \vspace{\baselineskip}
  
$\absolute{\alpha} = \ell'$, $\absolute{\beta} = \absolute{\gamma} = \rhs - j$\hfill$\stringstrut%
\str = %
\stringbox[1mm]{}{}{}%
\stringbox[\tmpd]{\mathrlap{%
\hspace{3mm}\mathrlap{\stringidx{\ i''}}%
\phantom{\stringbox[.75mm]{}{\beta}{}%
\stringbox[0mm]{}{\gamma}{}}%
\hspace{2mm}\mathrlap{\stringidx{j''}}
}\stringidx{i'}}{\alpha}{}%
\stringbox[5mm]{\stringidx{\ \ \ \ (i' + \ell')}}{}{}%
\stringbox[\tmpd]{\mathrlap{%
\hspace{3mm}\mathrlap{\stringidx{i}}%
\phantom{\stringbox[.75mm]{}{\beta}{}%
\stringbox[0mm]{}{\gamma}{}}%
\hspace{2mm}\mathrlap{\stringidx{j}}
}\stringidx{j'}}{\alpha}{}%
\stringbox[1mm]{\stringidx{\rhs}}{}{}$   
    
$\ell'' \geq \absolute{\beta}$, $\ell \geq \absolute{\beta}$\hfill$\stringstrut%
\mathrlap{%
\hspace{3mm}\tikzmark{m2}\stringbox[.75mm]{}{\beta}{}\phantom{\stringbox[0mm]{}{\gamma}{}}%
\hspace{2mm}\stringbox[.75mm]{}{\mathrlap{\gamma}\phantom{\beta}}{}\quad\tikzmark{m3}%
\begin{tikzpicture}[overlay, remember picture]
  \foreach \x in {2,3} {
    \path (m\x) ++(3pt,-0pt) node (m\x) {}; 
  }
  \draw (m2) edge[looseness=.6, out = 270, in = 270, -Latex] (m3);
\end{tikzpicture}%
}
\tikzmark{m1}\phantom{\stringbox[\tmpd]{\stringidx{i'}}{\alpha}{}%
\stringbox[5mm]{}{}{}}%
\mathrlap{%
\hspace{3mm}\tikzmark{m5}\stringbox[.75mm]{}{\beta}{}%
\phantom{\stringbox[0mm]{}{\gamma}{}}%
\hspace{2mm}\tikzmark{m6}\stringbox[.75mm]{}{\mathrlap{\gamma}\phantom{\beta}}{}%
\begin{tikzpicture}[overlay, remember picture]
  \foreach \x in {5,6} {
    \path (m\x) ++(3pt,-0pt) node (m\x) {}; 
  }
  \draw (m5) edge[out = 270, in = 270, -Latex] (m6);
\end{tikzpicture}%
}
\tikzmark{m4}\phantom{\stringbox[\tmpd]{\stringidx{j'}}{\alpha}{}%
\stringbox[1mm]{}{}{}}%
\begin{tikzpicture}[overlay, remember picture]
  \path (m1) ++(3pt,12pt) node (m1) {};
  \path (m4) ++(3pt,-2pt) node (m4) {};

  \draw[-Latex] (m1) to ++(0,-14pt) to[looseness = .7, out=270, in=270] (m4.center) to ++(0,11pt);
    
\end{tikzpicture}$

  \vspace{2\baselineskip}
  
  First, note that $\str[i'..i' + \ell') = \str[j'..\rhs)$ implies $\str[i..j) = \str[i''..j'')$. 
  From $\nss{i} = j$ follows that $\str[i..j) = \str[i''..j'')$ is a Lyndon word.
  Thus, due to \cref{lemma:lyndonnss} and $\nss{i''} \neq j''$ it holds $\nss{i''} > j''$, which implies $\str_{i''} \llex \str_{j''}$. 
  Let $\beta = \str[i''..i''+ \rhs - j) = \str[i..i + \rhs - j)$ and let $\gamma = \str[j''..i'+ \ell') = \str[j..\rhs)$.
  From $\str_{i''} \llex \str_{j''}$ follows $\beta \leqlex \gamma$, while $\str_i \glex \str_j$ implies $\beta \geqlex \gamma$.
  Thus it holds $\beta = \gamma$, and therefore $\rlce{i, j} \geq \absolute{\gamma} = \rhs - j$.
  This means that we can use the strategy from Case R1.
  
  {\textcolor{black!30!white}{\hrulefill}\\[-.5\baselineskip]}    
  
  \case{R3b} \textbf{\boldmath$\nss{i''} = j''$ and\\$(j'' + \ell'') < (i' + \ell')$\unboldmath:}%

\hfill$\stringstrut%
\str = %
\stringbox[1mm]{}{}{}%
\stringbox[\tmpd]{\mathrlap{%
\hspace{4mm}\mathrlap{\stringidx{i''}}%
\phantom{\stringbox[.2em]{}{\beta}{}}%
\hspace{3mm}\mathrlap{\stringidx{j''}}
}\stringidx{i'}}{\alpha}{}%
\stringbox[5mm]{\stringidx{\ \ \ (i' + \ell')}}{}{}%
\stringbox[\tmpd]{\mathrlap{%
\hspace{4mm}\mathrlap{\stringidx{i}}%
\phantom{\stringbox[.2em]{}{\beta}{}}%
\hspace{3mm}\mathrlap{\stringidx{j}}
}\stringidx{j'}}{\alpha}{}%
\stringbox[1mm]{\stringidx{\rhs}}{}{}$   
    
$\absolute{\alpha} = \ell'$, $\absolute{\beta} = \ell'' = \ell$\hfill$\stringstrut%
\mathrlap{%
\hspace{4mm}\tikzmark{m2}\stringbox[.2em]{}{\beta}{}%
\hspace{3mm}\tikzmark{m3}\stringbox[.2em]{}{\beta}{}%
\begin{tikzpicture}[overlay, remember picture]
  \foreach \x in {2,3} {
    \path (m\x) ++(3pt,-0pt) node (m\x) {}; 
  }
  \draw (m2) edge[out = 270, in = 270, -Latex] (m3);
\end{tikzpicture}%
}
\tikzmark{m1}\phantom{\stringbox[\tmpd]{\stringidx{i'}}{\alpha}{}%
\stringbox[5mm]{}{}{}}%
\mathrlap{%
\hspace{4mm}\tikzmark{m5}\stringbox[.2em]{}{\beta}{}%
\hspace{3mm}\tikzmark{m6}\stringbox[.2em]{}{\beta}{}%
\begin{tikzpicture}[overlay, remember picture]
  \foreach \x in {5,6} {
    \path (m\x) ++(3pt,-0pt) node (m\x) {}; 
  }
  \draw (m5) edge[out = 270, in = 270, -Latex] (m6);
\end{tikzpicture}%
}
\tikzmark{m4}\phantom{\stringbox[\tmpd]{\stringidx{j'}}{\alpha}{}%
\stringbox[1mm]{}{}{}}%
\begin{tikzpicture}[overlay, remember picture]
  \path (m1) ++(3pt,12pt) node (m1) {};
  \path (m4) ++(3pt,0pt) node (m4) {};

  \draw[-Latex] (m1) to ++(0,-12pt) to[looseness = .7, out=270, in=270] (m4.center) to ++(0,9pt);
    
\end{tikzpicture}$

  \vspace{1.75\baselineskip}
  
  Due to $\ell'' = \rlce{i'', j''}$, there is a shared prefix $\beta = \str[i''..i'' + \ell'') = \str[j''..j'' + \ell'')$ between $\str_{i''}$ and $\str_{j''}$, and the first mismatch between the two suffixes is $\str[i'' + \ell''] \neq \str[j'' + \ell'']$. 
  Because of $(j'' + \ell'') < (i' + \ell')$, both the shared prefix and the mismatch are contained in $\str[i'..i' + \ell')$ (i.e.\ in the first occurrence of $\alpha$). 
  If we consider the substring $\str[j'..j' + \ell')$ instead (i.e.\ the second occurrence of $\alpha$), then $\str_i$ and $\str_j$ clearly also share the prefix $\beta = \str[i..i + \ell'') = \str[j..j + \ell'')$, with the first mismatch occurring at $\str[i + \ell''] \neq \str[j + \ell'']$.
  Thus it holds $\ell = \ell''$.
  Due to $\nss{i''} = j''$ and our order of R\=/LCE computations, we have already computed $\ell''$. 
  Therefore, we can simply assign $\ell \gets \ell''$ and continue without changing $i'$, $j'$, $\ell'$, and $\rhs$.
      
  {\textcolor{black!30!white}{\hrulefill}\\[-.5\baselineskip]}      
  
  \case{R3c} \textbf{\boldmath$\nss{i''} = j''$ and\\$(j'' + \ell'') \geq (i' + \ell')$\unboldmath:}

\hfill$\stringstrut%
\str = %
\stringbox[1mm]{}{}{}%
\stringbox[\tmpd]{\mathrlap{%
\hspace{3mm}\mathrlap{\stringidx{\ i''}}%
\phantom{\stringbox[.75mm]{}{\beta}{}%
\stringbox[0mm]{}{\gamma}{}}%
\hspace{2mm}\mathrlap{\stringidx{j''}}
}\stringidx{i'}}{\alpha}{}%
\stringbox[5mm]{\stringidx{\ \ \ \ (i' + \ell')}}{}{}%
\stringbox[\tmpd]{\mathrlap{%
\hspace{3mm}\mathrlap{\stringidx{i}}%
\phantom{\stringbox[.75mm]{}{\beta}{}%
\stringbox[0mm]{}{\gamma}{}}%
\hspace{2mm}\mathrlap{\stringidx{j}}
}\stringidx{j'}}{\alpha}{}%
\stringbox[1mm]{\stringidx{\rhs}}{}{}$   
    
\smash{$\absolute{\alpha} = \ell'$, $\absolute{\beta} = \rhs - j$, $\absolute{\beta\gamma} = \ell''$}\hfill$\stringstrut%
\mathrlap{%
\hspace{3mm}\tikzmark{m2}\stringbox[.75mm]{}{\beta}{}\stringbox[0mm]{}{\gamma}{}%
\hspace{2mm}\tikzmark{m3}\stringbox[.75mm]{}{\beta}{}\stringbox[0mm]{}{\gamma}{}%
\begin{tikzpicture}[overlay, remember picture]
  \foreach \x in {2,3} {
    \path (m\x) ++(3pt,-0pt) node (m\x) {}; 
  }
  \draw (m2) edge[out = 270, in = 270, -Latex] (m3);
\end{tikzpicture}%
}
\tikzmark{m1}\phantom{\stringbox[\tmpd]{\stringidx{i'}}{\alpha}{}%
\stringbox[5mm]{}{}{}}%
\mathrlap{%
\hspace{3mm}\tikzmark{m5}\stringbox[.75mm]{}{\beta}{}%
\phantom{\stringbox[0mm]{}{\gamma}{}}%
\hspace{2mm}\tikzmark{m6}\stringbox[.75mm]{}{\beta}{}%
\begin{tikzpicture}[overlay, remember picture]
  \foreach \x in {5,6} {
    \path (m\x) ++(3pt,-0pt) node (m\x) {}; 
  }
  \draw (m5) edge[out = 270, in = 270, -Latex] (m6);
\end{tikzpicture}%
}
\tikzmark{m4}\phantom{\stringbox[\tmpd]{\stringidx{j'}}{\alpha}{}%
\stringbox[1mm]{}{}{}}%
\begin{tikzpicture}[overlay, remember picture]
  \path (m1) ++(3pt,12pt) node (m1) {};
  \path (m4) ++(3pt,-2pt) node (m4) {};

  \draw[-Latex] (m1) to ++(0,-14pt) to[looseness = .7, out=270, in=270] (m4.center) to ++(0,11pt);
    
\end{tikzpicture}$

  $\ell \geq \absolute{\beta}$

  \vspace{\baselineskip}
  
  This situation is similar to Case R3b. 
  There is a shared prefix $\beta = {\str[i''..i'' + \rhs - j)} = \str[j''..i' + \ell')$ between the suffixes $\str_{i''}$ and $\str_{j''}$. 
  They may share an even longer prefix $\beta\gamma$, but only the first $\absolute{\beta} = \rhs - j$ symbols of their LCP are contained in $\str[i'..i' + \ell')$ (i.e.\ in the first occurrence of $\alpha$). 
  If we consider the substring $\str[j'..j' + \ell')$ instead (i.e.\ the second occurrence of $\alpha$), then $\str_i$ and $\str_j$ clearly also share at least the prefix $\beta = \str[i..i + \rhs - j) = \str[j..\rhs)$.
  Thus it holds $\rhs - j \leq \ell$, and we can use the strategy from Case R1.  
\end{itembox}

We have shown how to compute $\ell$ without charging any index twice.
It follows that the total number of character comparisons for all R\=/LCEs is $\orderof{n}$.

\subparagraph*{A Simple Algorithm for R\=/LCEs.} 
While the detailed differentiation between the six subcases helps to show the correctness of our approach, it can be implemented in a significantly simpler way (see \cref{alg:rlce}).
At all times, we keep track of $j'$, $\rhs$ and the distance $d = j' - i'$ (line~\ref{alg:rlce:init}). 
We consider the indices $j \in [2, n]$ in increasing order (line~\ref{alg:rlce:outerloop}). 
For each index $j$, we then consider the indices $i$ with $\nss{i} = j$ in decreasing order (line~\ref{alg:rlce:innerloop}).
Each time we want to compute an R\=/LCE value $\ell = \rlce{i, j}$, we first check whether Case R3b applies~(line~\ref{alg:rlce:case3a}).
If it does, then we simply copy the previously computed R\=/LCE value $\rlce{i - d, j - d}$ (line \ref{alg:rlce:copylce}). 
Otherwise, we either compute the LCE naively (if $j \geq \rhs$), or we have to apply the strategy from Case R1 (since all other cases except for Case R2b use this strategy; in Case R2b it holds $\rhs - j = \ell$, which means that it can also be solved with the strategy from Case R1). 
If $j \geq \rhs$ then in lines \ref{alg:rlce:computelce1}--\ref{alg:rlce:computelce2} we have $k = 0$, and thus we naively compute $\rlce{i, j}$ by scanning. 
If however $j < \rhs$, then we have $k = \rhs - j$, and we skip $k$ character comparisons. 
In any case, we update the values $j'$, $\rhs$, and $d$ accordingly (line \ref{alg:rlce:updatevars}).

\begin{algorithm}
\newcommand{\naiverlce}[1]{\textsc{naive-scan-}\rlce{#1}}
\newcommand{\prevj}{j_{\text{prev}}}
\newcommand{\knownlce}{k}
\newcommand{\addlce}{a}
\let\oldstrut\strut
\renewcommand{\strut}{\vphantom{\rhs}\oldstrut}
\begin{algorithmic}[1]
  \Require{String $\str$ of length $n$ and its NSS array \textsf{nss}.}
  \Ensure{R\=/LCE value $\rlce{i, \nss{i}}$ for each index $i \in [1, n]$ with $\nss{i} \neq n + 1$.}
  
  \vspace{.25\baselineskip} 

  \State $j' \gets 0;\ \ \rhs \gets 1;\ \ d \gets 0$\label{alg:rlce:init}
  
  \vspace{.25\baselineskip} 
    
  \For{$j \in [2, n]$ in increasing order}\label{alg:rlce:outerloop}
  \vspace{.25\baselineskip} 
  \For{$i$ \textbf{with} $\nss{i} = j \neq n + 1$ in decreasing order}\label{alg:rlce:innerloop}
  \vspace{.5\baselineskip}  
  \If{$\left(\parbox[m]{4.2cm}{%
  \vspace{-\baselineskip}
  \begin{alignat*}{2}
  \strut &i, j \in (j', \rhs)\\[-.25\baselineskip]
  \strut\land\ & \nss{i - d} = j - d\\[-.25\baselineskip]
  \strut\land\ & j + \rlce{i - d, j - d} < \rhs
  \end{alignat*}
  \vspace{-1.3\baselineskip}}\right)$}\label{alg:rlce:case3a}
  \vspace{.35\baselineskip}
  \State $\rlce{i, j} \gets \rlce{i - d, j - d}$\label{alg:rlce:copylce}
  \vspace{.5\baselineskip}  
  \Else
  \State $\strut\knownlce \gets \max(\rhs, j) - j$ \label{alg:rlce:computelce1}
  \State $\strut\rlce{i, j} \gets k + \naiverlce{i + \knownlce, j + \knownlce}$ \label{alg:rlce:computelce2}
  \State $\strut\mathrlap{j'}\phantom{\rhs} \gets j;\ \ \rhs \gets j + \rlce{i, j};\ \ \mathrlap{d}\phantom{\rhs} \gets j - i$ \label{alg:rlce:updatevars}
  \EndIf
  \EndFor
  \EndFor
\end{algorithmic}
\caption{Compute all R\=/LCEs.}
\label{alg:rlce}
\end{algorithm}

The correctness of the algorithm follows from the description of Cases 1--3.
Since for each left index $i$ we have to store at most one R\=/LCE, we can simply maintain the LCEs in a length-$n$ array, where the $i$-th entry is $\rlce{i, \nss{i}}$.
This way, we use linear space and can access the R\=/LCE that is required in line~\ref{alg:rlce:copylce} in constant time.
Apart from the at most $n$ character comparisons that we charge to the indices, we only need a constant number of additional primitive operations per computed R\=/LCE. 
The order of iteration can be realized by first generating all $(i, \nss{i})$-pairs, and then using a linear time radix sorter to sort the pairs in increasing order of their second component and decreasing order of their first component.
We have shown:

\begin{lemma}\label{lemma:linearrlce}
	Given a string of length $n$ and its NSS array \textnormal{\textsf{nss}}, we can compute $\rlce{i, \nss{i}}$ for all indices $i \in [1, n]$ with $\nss{i} \neq n + 1$ in $\orderof{n}$ time and space.
\end{lemma}

\subsection{Computing the L\=/LCE Values}
\label{sec:llce}

Our solution for the L\=/LCEs is similar to the one for R\=/LCEs, but differs in subtle details.
We generally compute $\ell = \llce{i, j}$ by simultaneously scanning the prefixes $\str[1..i]$ and $\str[1..j]$ from right to left until we find the first mismatch.
This takes $\ell + 1$ character comparisons, of which we charge $\ell$ comparisons to the interval $(i - \ell, i]$. 
As before, if some lower bound $k \leq \ell$ is known then we skip $k$ character comparisons. 
In this case, we compute the L\=/LCE as $\ell = k + \llce{i - k, j - k}$, and charge $\ell - k$ comparisons to the interval $(i - \ell, i - k]$.

Again, we will show how to compute all values $\llce{i, \nss{i}}$ with $i \in [1, n]$ and $\nss{i} \neq n + 1$ such that each index gets charged at most once. 
In contrast to the more complex R\=/LCE iteration order, we can simply compute the L\=/LCE values in \emph{decreasing} order of~$i$.
Thus, when we want to compute $\ell = \llce{i, j}$ with $j = \nss{i} \neq n + 1$, we have already considered the indices $I = \{ x \mid x \in (i, n] \land \nss{x} \neq n + 1 \}$ as left indices of L\=/LCE computations. The leftmost text position that we have already inspected so far at this point is $\lhs = \min_{x \in I}(x - \llce{x, \nss{x}})$ if $I \neq \emptyset$, or $\lhs = n$ otherwise.
%
%
Due to our charging method, we have not charged any index from the interval $[1, \lhs]$ yet. Thus, we only have to show how to compute $\ell$ without charging indices from $(\lhs, n]$.
Let $\ell' = \llce{i', j'}$ be the most recently computed L\=/LCE that satisfies $i' - \ell' = \lhs$.
If $i \leq \lhs$ then we compute $\ell$ naively and charge the character comparisons to the interval $(i - \ell, i]$ (thus only charging previously uncharged indices). If however $i > \lhs$, then our strategy is more complicated. Before explaining it in detail, we show three important properties that hold in the present situation.

\vspace{.5\baselineskip}

\begin{figure}

\phantom{a}

\vspace{\baselineskip}

\subcaptionbox{\label{fig:llce:sup1}}%
{$\str{=}%
\stringbox[0pt]{}{\ \stringidx{\lhs}\enskip\stringidx{i}\quad\enskip}{}
\tikzmark{1}%
\stringbox[.4cm]{\stringidx[*]{\enskip\ i' - (j' - i')}}{\beta}{}
\tikzmark{e1}\tikzmark{2}%
\stringbox[.4cm]{\stringidx[*]{i'}}{\beta}{}
\tikzmark{e2}%
\stringbox[2pt]{\stringidx[*]{j'}}{\ }{}%
\begin{tikzpicture}[overlay, remember picture]
	\path (1) ++(3pt,-4pt) node (1) {} ++(0,-6pt) node (i1) {};
	\foreach \x in {2} {
		\path (\x) ++(4.5pt,-10pt) node (\x) {} ++(0,-0pt) node (i\x) {};
	}
	\foreach \x in {1,2} {
		\path (e\x) ++(3pt,-4pt) node (e\x) {} ++(0,-6pt) node (ie\x) {};
	}
	\draw[-Latex, dashed] (1.center) to (i1.center) to[out = 270, in = 270, looseness = .7] (ie1.center) to (e1.center);
	\draw[-Latex] (2.center) to (i2.center) to[out = 270, in = 270, looseness = .7] (ie2.center) to (e2.center);
\end{tikzpicture}$}\hfill%
\subcaptionbox{\label{fig:llce:sup2}}{$\stringstrut%
\str{=}%
\stringbox[1mm]{}{}{\stringidx{\lhs}}%
\stringbox[1.25mm]{}{\gamma}{}%
\tikzmark{m1}%
\stringbox[2mm]{\stringidx{i}}{\beta}{\stringidx{i'}}%
\tikzmark{m2}%
\stringbox[2mm]{}{}{}%
\stringbox[1.25mm]{}{\gamma}{}%
\stringbox[2mm]{\stringidx{(j' - i' + i)\qquad}}{\beta}{\stringidx{j'}}%
\tikzmark{m3}%
\stringbox[1mm]{}{\ \stringidx{j}}{}%
\begin{tikzpicture}[overlay, remember picture]
  \path (m1) ++(3pt,-0pt) node (m1) {}; 
  \path (m2) ++(-3pt,-0pt) node (m2) {}; 
  \path (m3) ++(-3pt,-0pt) node (m3) {};
  \path (m3) ++(13pt,-0pt) node (m4) {};
  \draw (m2) edge[looseness=.5, out = 270, in = 270, -Latex] (m3);
  \draw (m1) edge[looseness=.5, out = 270, in = 270, -Latex] (m4);
\end{tikzpicture}%
$}\hfill%
\subcaptionbox{\label{fig:llce:sup3}}{$\str{=}\stringbox[3pt]{}{%
  \stringidx{\lhs}%
  \enskip%
  \tikzmark{s1}\stringidx{i}\quad\enskip%
  \tikzmark{t1}\stringidx{j}\quad%
\tikzmark{s2}\stringidx{i'}%
\mathllap{\stringbox[5.1mm]{}{\alpha}{}\hspace{-1pt}}%
\quad\quad%
\tikzmark{s3}\stringidx{\ i''}\quad\enskip%
\tikzmark{t3}\stringidx{\ j''}\quad%
\tikzmark{t2}\stringidx{j'}%
\mathllap{\stringbox[5.1mm]{}{\alpha}{}\hspace{-1pt}}%
}{}%
\begin{tikzpicture}[overlay, remember picture]
	\foreach \x in {1,2} {
		\path (s\x) ++(1pt, 0) node (s\x) {};
		\path (t\x) ++(1pt, 0) node (t\x) {} ++(0,-10pt) node (m\x) {};
		\draw[-Latex] (s\x) to ++(0, -10pt) to[out=270, in=270, looseness=.5] (m\x.center) to (t\x);
	}
	\foreach \x in {3} {
		\path (s\x) ++(1pt, 0) node (s\x) {};
		\path (t\x) ++(1pt, 0) node (t\x) {} ++(0,-10pt) node (m\x) {};
		\draw[-Latex, dashed] (s\x) to ++(0, -10pt) to[out=270, in=270, looseness=.5] (m\x.center) to (t\x);
	}
\end{tikzpicture}$}
\caption{Illustration of the proofs of the three properties in \cref{sec:llce}.}
\label{fig:llce:supplement}
\end{figure}
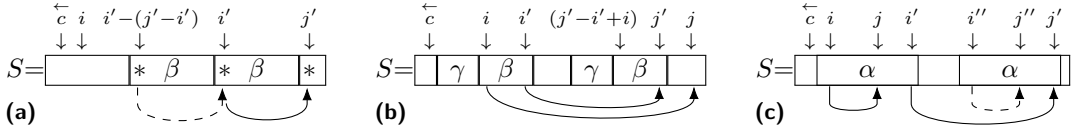

First, we show that ${i \geq i' - (j' - i')}$. Assume the opposite (as visualized in \cref{fig:llce:sup1}), then from ${\lhs = i' - \ell' < i}$ follows ${\ell' > j' - i'}$. Thus, \cref{lemma:lyndonchainleft} implies ${\nss{i' - (j' - i')} = i'}$ (dashed edge) and ${\llce{i' - (j' - i'), i'}} = \ell' - (j' - i')$. 
Due to our order of computation and $i < i' - (j' - i')$ we must have already computed this L-LCE.
However, it holds $i' - (j' - i') - \llce{i' - (j' - i'), i'} = 
\rhs$, which contradicts the fact that $\ell' = \llce{i', j'}$ is the \emph{most recently} computed L-LCE with $i' - \ell' = \lhs$.

Next, we show that $j \leq i'$.
First, note that $j \notin (i', j')$, since due to $i < i'$ we would otherwise contradict \cref{lemma:nonintersecting}.
Thus we only have to show ${j < j'}$.
Assume for this purpose that $j \geq j'$ (as visualized in \cref{fig:llce:sup2}).
From ${j' - i' + i} \in (i, \nss{i})$ and \cref{def:xss} follows ${\str_i \llex \str_{j' - i' + i}}$. 
Because of $\llce{i', j'} > (i' - i)$ it holds $\str[i..i'] = \str[j' - i' + i..j'] (= \beta)$. Thus ${\str_i \llex \str_{j' - i' + i}}$ implies $\str_{i'} \llex \str_{j'}$, which contradicts the fact that $\nss{i'} = j'$. 

Lastly, let $d = j' - i'$, $i'' = i + d$, and $j'' = j + d$ (as visualized in \cref{fig:llce:sup3}).
Now we show that $\nss{i''} = j''$ (dashed edge in the figure).
Because of $\alpha = \str(\lhs..i'] = \str(j' - \ell'..j']$ it holds $\str[i..j) = \str[i''..j'')$.
From $\nss{i} = j$ and \cref{lemma:lyndonnss} follows that $\str[i''..j'')$ is a Lyndon word, and thus $\nss{i''} \geq j''$.
We have already shown that $i \geq i' - (j' - i')$, which implies $i'' \geq i'$.
Due to $\nss{i'} = j'$ and $i'' \in [i', j')$ it follows from \cref{lemma:nonintersecting} that $\nss{i''} \leq j'$.
Now assume $\nss{i''} \in (j'', j']$, then $\str[i''..\nss{i''}) = \str[i..j + (\nss{i''} - j''))$ is a Lyndon word, which contradicts the fact that $\str[i..j)$ is the longest Lyndon word starting at position $i$.
Thus, we have ruled out all possible values of $\nss{i''}$ except for $j''$.

\vspace{.5\baselineskip}

Now we show how to compute $\ell$. We keep using the definition of $i''$ and $j''$ from the previous paragraph. Furthermore, let $\ell'' = \llce{i'', j''}$. There are two possible cases.


  \renewcommand{\tmpd}{7.5mm}
  
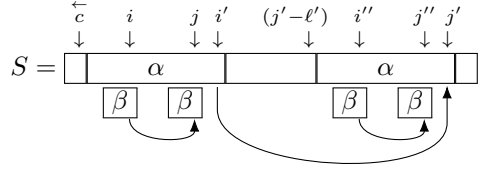
  \begin{itembox}%
  \case{L1} \bm{(i'' - \ell'') > (j' - \ell')}\textbf{.}
  
  \vspace{.5\baselineskip}
  
  \hfill$\stringstrut%
\str = %
\stringbox[1mm]{}{}{\stringidx{\lhs}}%
\stringbox[\tmpd]{}{\alpha}{\stringidx{\ i'}%
\mathllap{
\mathllap{\stringidx{i}}
\phantom{\stringbox[.2em]{}{\beta}{}\hspace{4mm}}
\mathllap{\stringidx{j}}\hspace{3mm}%
}}%
\stringbox[5.5mm]{}{}{\stringidx{(j' - \ell')\ \ \ }}%
\stringbox[\tmpd]{}{\alpha}{\stringidx{\ \,j'}%
\mathllap{
\mathllap{\stringidx{\ i''}}
\phantom{\stringbox[.2em]{}{\beta}{}\hspace{4mm}}
\mathllap{\stringidx{j''}}\hspace{3mm}%
}}%
\stringbox[1mm]{}{}{}$   
    
$\ell' = \absolute{\alpha}$, $\ell = \ell'' = \absolute{\beta}$\hfill%
$\stringstrut%
\mathllap{%
\stringbox[.2em]{}{\beta}{}\tikzmark{m2}\hspace{4mm}%
\stringbox[.2em]{}{\beta}{}\tikzmark{m3}\hspace{3mm}%
\begin{tikzpicture}[overlay, remember picture]
  \foreach \x in {2,3} {
    \path (m\x) ++(-3pt,-0pt) node (m\x) {}; 
  }
  \draw (m2) edge[out = 270, in = 270, -Latex] (m3);
\end{tikzpicture}}%
\tikzmark{m1}%
\phantom{\stringbox[5.5mm]{}{}{}}%
\phantom{\stringbox[\tmpd]{}{\alpha}{}}%
\mathllap{%
\stringbox[.2em]{}{\beta}{}\tikzmark{m5}\hspace{4mm}%
\stringbox[.2em]{}{\beta}{}\tikzmark{m6}\hspace{3mm}%
\begin{tikzpicture}[overlay, remember picture]
  \foreach \x in {5,6} {
    \path (m\x) ++(-3pt,-0pt) node (m\x) {}; 
  }
  \draw (m5) edge[out = 270, in = 270, -Latex] (m6);
\end{tikzpicture}}%
\tikzmark{m4}%
\phantom{\stringbox[1mm]{}{}{}}%
\begin{tikzpicture}[overlay, remember picture]
  \path (m1) ++(-3pt,12pt) node (m1) {};
  \path (m4) ++(-3pt,-5pt) node (m4) {};

  \draw[-Latex] (m1) to ++(0,-12pt) to[looseness = .7, out=270, in=270] (m4.center) to ++(0,14pt);
    
\end{tikzpicture}$

  \vspace{2.25\baselineskip}

  Due to $\ell'' = \llce{i'', j''}$, the prefixes $\str[1..i'']$ and $\str[1..j'']$ share the suffix $\beta = \str(i'' - \ell''..i''] = \str(j'' - \ell''..j'']$, and the first (from the right) mismatch between these prefixes is $\str[i'' - \ell''] \neq \str[j'' - \ell'']$. 
  Both the shared suffix and the mismatch are contained in $\str(j' - \ell'..j']$ (i.e.\ in the right occurrence of $\alpha$). If we consider the substring $\str(\lhs..i']$ instead (i.e.\ the left occurrence of $\alpha$), then $\str[1..i]$ and $\str[1..j]$ clearly also share the suffix $\beta = \str(i - \ell''..i] = \str(j - \ell''..j]$, with the first mismatch occurring at $\str[i - \ell''] \neq \str[j'' - \ell]$. Thus it holds $\ell = \ell''$. Due to $\nss{i''} = j''$ and our order of L-LCE computations, we have already computed $\ell''$. Therefore, we can simply assign $\ell \gets \ell''$ and continue without changing $i'$, $j'$, $\ell'$, and $\lhs$.
  
  (Note that possibly $i'' \neq i' \land j'' = j'$. We provide a sketch in \cref{appendix}, \cref{fig:NL1}.)
  \end{itembox}
  
  \begin{itembox}
  \case{L2} \bm{(i'' - \ell'') \leq (j' - \ell')}\textbf{.}
  
  \vspace{.5\baselineskip}
  
  \hfill$\stringstrut%
\str = %
\stringbox[1mm]{}{}{\stringidx{\lhs}}%
\stringbox[\tmpd]{}{\alpha}{\stringidx{\ i'}%
\mathllap{
\mathllap{\stringidx{i}}
\phantom{\stringbox[.28em]{}{\beta}{}\hspace{5mm}}
\mathllap{\stringidx{j}}\hspace{3mm}%
}}%
\stringbox[5.5mm]{}{}{\stringidx{(j' - \ell')\ \ \ }}%
\stringbox[\tmpd]{}{\alpha}{\stringidx{\ \,j'}%
\mathllap{
\mathllap{\stringidx{\ i''}}
\phantom{\stringbox[.28em]{}{\beta}{}\hspace{5mm}}
\mathllap{\stringidx{j''}}\hspace{3mm}%
}}%
\stringbox[1mm]{}{}{}$   
    
$\ell' = \absolute{\alpha}$, $\ell'' = \absolute{\beta\gamma}$, $\ell \geq \absolute{\beta}$%
\hfill%
$\stringstrut%
\mathllap{%
\stringbox[.28em]{}{\beta}{}\tikzmark{m2}\hspace{5mm}%
\stringbox[.28em]{}{\beta}{}\tikzmark{m3}\hspace{3mm}%
\begin{tikzpicture}[overlay, remember picture]
  \foreach \x in {2,3} {
    \path (m\x) ++(-3pt,-0pt) node (m\x) {}; 
  }
  \draw (m2) edge[out = 270, in = 270, -Latex] (m3);
\end{tikzpicture}}%
\tikzmark{m1}%
\phantom{\stringbox[5.5mm]{}{}{}}%
\phantom{\stringbox[\tmpd]{}{\alpha}{}}%
\mathllap{%
\mathllap{\stringbox[0mm]{}{\gamma}{}}%
\stringbox[.28em]{}{\beta}{}\tikzmark{m5}\hspace{5mm}%
\mathllap{\stringbox[0mm]{}{\gamma}{}}%
\stringbox[.28em]{}{\beta}{}\tikzmark{m6}\hspace{3mm}%
\begin{tikzpicture}[overlay, remember picture]
  \foreach \x in {5,6} {
    \path (m\x) ++(-3pt,-0pt) node (m\x) {}; 
  }
  \draw (m5) edge[out = 270, in = 270, -Latex] (m6);
\end{tikzpicture}}%
\tikzmark{m4}%
\phantom{\stringbox[1mm]{}{}{}}%
\begin{tikzpicture}[overlay, remember picture]
  \path (m1) ++(-3pt,12pt) node (m1) {};
  \path (m4) ++(-3pt,-5pt) node (m4) {};

  \draw[-Latex] (m1) to ++(0,-12pt) to[looseness = .7, out=270, in=270] (m4.center) to ++(0,14pt);
    
\end{tikzpicture}$

  \vspace{2.25\baselineskip}

  This situation is similar to Case L1. There is a shared suffix $\beta = \str(j' - \ell'..i''] = \str(j'' - (i - \lhs)..j'']$ between the prefixes $\str[1..i'']$ and $\str[1..j'']$.
  They may share an even longer suffix $\gamma\beta$, but only the rightmost $\absolute{\beta} = i' - \lhs$ symbols of this suffix are contained in $\str(j' - \ell'..j']$ (i.e.\ in the right occurrence of $\alpha$). If we consider the substring $\str(\lhs..i']$ instead (i.e.\ the left occurrence of $\alpha$), then $\str[1..i]$ and $\str[1..j]$ clearly also share the suffix $\beta = \str(\lhs..i] = \str(j - (i - \lhs)..j]$. Thus it holds $i - \lhs \leq \ell$, and we can skip the first $i - \lhs$ character comparisons by computing the LCE as $\ell = (i - \lhs) + \llce{\lhs, j + \lhs - i}$. We charge $\ell - (i - \lhs)$ character comparisons to the previously uncharged interval $(i - \ell, \lhs]$, and continue with $i' \gets i$, $j' \gets j$, $\ell' \gets \ell$, and $\lhs \gets i - \ell$.
  
  (Note that possibly $i'' \neq i' \land j'' = j'$ or even $i'' = i' \land j'' = j'$. We provide schematic drawings in \cref{appendix}, \cref{fig:NL2a,fig:NL2b}.)
  \end{itembox}
  
We have shown how to compute $\ell$ without charging any index twice.
It follows that the total number of character comparisons for all LCEs is $\orderof{n}$.
For completeness, we outline a simple implementation of our approach in \cref{alg:llce}.
Lines~\ref{alg:llce:case1}--\ref{alg:llce:copylce} correspond to Case L1. If $i \leq \lhs$, then lines~\ref{alg:llce:computelce1}--\ref{alg:llce:updatevars} compute the LCE naively. Otherwise, they correspond to Case L2.

\begin{algorithm}
\newcommand{\naivellce}[1]{\textsc{naive-scan-}\llce{#1}}
\newcommand{\prevj}{j_{\text{prev}}}
\newcommand{\knownlce}{k}
\newcommand{\addlce}{a}
\let\oldstrut\strut
\renewcommand{\strut}{\vphantom{\rhs}\oldstrut}
\begin{algorithmic}[1]
  \Require{String $\str$ of length $n$ and its NSS array \textsf{nss}.}
  \Ensure{L\=/LCE value $\llce{i, \nss{i}}$ for each index $i \in [1, n]$ with $\nss{i} \neq n + 1$.}
  
  \vspace{.25\baselineskip} 

  \State $i' \gets 0;\ \ \lhs \gets n;\ \ d \gets 0$\label{alg:llce:init}
  
  \vspace{.25\baselineskip} 
    
  \For{$i \in [1, n]$ \textbf{with} $\nss{i} \neq n + 1$ in decreasing order}\label{alg:llce:loop}
  \vspace{.25\baselineskip} 
  
  \State $j \gets \nss{i}$  
  \vspace{.5\baselineskip}  
  
  \If{$i \in (\lhs, i') \land i - \llce{i + d, j + d} > \lhs$}\label{alg:llce:case1}
  \vspace{.35\baselineskip}
  \State $\llce{i, j} \gets \llce{i + d, j + d}$\label{alg:llce:copylce}
  \vspace{.5\baselineskip}  
  \Else
  \State $\strut\knownlce \gets i - \min(\lhs, i)$ \label{alg:llce:computelce1}
  \State $\strut\llce{i, j} \gets k + \naivellce{i - \knownlce, j - \knownlce}$ \label{alg:llce:computelce2}
  \State $\strut\mathrlap{i'}\phantom{\lhs} \gets i;\ \ \lhs \gets i - \llce{i, j};\ \ \mathrlap{d}\phantom{\rhs} \gets j - i$ \label{alg:llce:updatevars}
  \EndIf
  \EndFor
\end{algorithmic}
\caption{Compute all L\=/LCEs.}
\label{alg:llce}
\end{algorithm}

\begin{lemma}\label{lemma:linearllce}
	Given a string of length $n$ and its NSS array \textnormal{\textsf{nss}}, we can compute $\llce{i, \nss{i}}$ for all indices $i \in [1, n]$ with $\nss{i} \neq n + 1$ in $\orderof{n}$ time and space.
\end{lemma}

\begin{corollary}
	Given a string of length $n$ over a general ordered alphabet, we can find all runs in the string in $\orderof{n}$ time and space.
	\begin{proof}
		Computing the increasing runs takes $\orderof{n}$ time and space due to \cref{lemma:requiredlces,lemma:linearrlce,lemma:linearllce}. For decreasing runs, we only have to reverse the order of the alphabet and rerun the algorithm.
	\end{proof}
\end{corollary}

\section{Practical Implementation}
\label{sec:practical}

We implemented our algorithm for the runs computation in C++17 and evaluated it by computing all runs on texts from the natural, real repetitive, and artificial repetitive text collections of the Pizza-Chili corpus\footnote{\url{http://pizzachili.dcc.uchile.cl/texts.html}, \url{http://pizzachili.dcc.uchile.cl/repcorpus.html}}. Additionally, we used the binary run-rich strings proposed by Matsubara et al. \cite{Matsubara2009} as input. \cref{table:results} shows the throughput that we achieve, i.e.\ the number of input bytes (or equivalently input symbols) that we process per second. On the string \texttt{tm29} we achieve the highest throughput of $15.6$ MiB/s. The lowest throughput of $8.8$ MiB/s occurs on the text \texttt{dna}.
Generally, we perform better for run-rich strings.

\begin{table}
\newcommand{\intxt}[2][]{\smash{\rotatebox{70}{\rlap{\shortstack{\bfseries\texttt{#2}\\{\scriptsize #1 MiB}}}}}}
\caption{Throughput achieved by our runs algorithm using an AMD EPYC 7452 processor. We repeated each experiment five times and use the median throughput as the final result. All numbers are truncated to one decimal place.}
\vspace{2.0\baselineskip}
\begin{tabular}{c|c|cccccc|cc|ccc}
        \intxt[$n$ in]{\textnormal{\bfseries Text}} & \intxt[1077]{\textbf{\boldmath$t_{49}$\unboldmath \cite{Matsubara2009}}} & \intxt[201]{sources} & \intxt[53]{pitches} & \intxt[1024]{proteins} & \intxt[385]{dna} & \intxt[1024]{english} & \intxt[282]{xml} & \intxt[107]{ecoli} & \intxt[439]{cere} & \intxt[255]{fib41} & \intxt[206]{rs.13} & \intxt[256]{tm29} \\ \hline
runs/$100n$ & 94.4                                  & 4.7             & 11.7            & 7.0              & 25.3        & 2.4             & 3.4         & 24.4          & 23.6                     & 76.3          & 92.7          & 83.3         \\
MiB/s    & 15.0                                  & 11.4            & 11.0            & 10.9             & 8.8         & 10.5            & 12.8        & 9.0           & 9.2                    & 15.4          & 15.1          & 15.6        
\end{tabular}
\label{table:results}
\end{table}

Lastly, it is noteworthy that our new method of LCE computation leads to a remarkably simple implementation of the runs algorithm. 
In fact, the entire implementation \emph{including the computation of the NSS array} needs only 250 lines of code.
We achieve this by interleaving the computation of the R-LCEs with the computation of the NSS array, which also improves the practical performance.
For technical details we refer to the source code, which is publicly available on GitHub\footnote{\url{https://github.com/jonas-ellert/linear-time-runs/}}.

\section{Conclusion and Open Questions}
\label{sec:conclusion}

We have shown the first linear time algorithm for computing all runs on a general ordered alphabet.
The algorithm is also very fast in practice and remarkably easy to implement.
It is an open question whether our techniques could be used for the computation of runs on tries, where the best known algorithms require superlinear time even for linearly-sortable alphabets (see e.g. \cite{Sugahara2019}).

\bibliography{literature}

\vfill
\appendix

\section{Supplementary Material}
\label{appendix}

\begin{figure}[h!]
\subcaptionbox{Case L1 with $i'' \neq i'$ and $j'' = j'$.\label{fig:NL1}}{\parbox{.47\textwidth}{
\flushleft$\strut$\\$\strut$\\$\stringstrut%
\str = %
\stringbox[1mm]{}{}{\stringidx{\lhs}}%
\stringbox[\tmpd]{}{\mathllap{\stringidx{i}\enskip}\alpha}{\stringidx{j = i'}}%
\stringbox[5mm]{}{}{\stringidx{(j' - \ell')\ \ \ }}%
\stringbox[\tmpd]{}{\mathllap{\stringidx{\ i''}\enskip}\alpha}{\stringidx{j'' = j'}}%
\stringbox[1mm]{}{}{}$

$\stringstrut%
\phantom{\str = \stringbox[1mm]{}{}{\stringidx{}}}%
\stringstrut%
\mathrlap{\ \,\stringbox[.5mm]{}{\beta}{}}
\phantom{\stringbox[\tmpd]{}{\alpha}{\stringidx{}}}%
\tikzmark{m1}\mathllap{\stringbox[.5mm]{}{\beta}{}}
\phantom{\stringbox[5mm]{}{}{\stringidx{}}}%
\mathrlap{\ \,\stringbox[.5mm]{}{\beta}{}}
\phantom{\stringbox[\tmpd]{}{\alpha}{\stringidx{}}}%
\tikzmark{m2}\mathllap{\stringbox[.5mm]{}{\beta}{}}
\phantom{\stringbox[1mm]{}{}{}}
\begin{tikzpicture}[overlay, remember picture]
  \foreach \x in {1,2} {
     \path
     (m\x) ++(-2pt,0) node (m\x) {}
     (m\x) ++(-12.5mm,0) node (e\x) {} 
     (m\x) ++(0,-10pt) node (md\x) {}
     (e\x) ++(0,-10pt) node (ed\x) {};
  }
  \draw[-Latex] (e1) to (ed1.center) to[in=270, out=270, looseness=.6] (md1.center) to (m1);
  \draw[-Latex] (e2) to (ed2.center) to[in=270, out=270, looseness=.6] (md2.center) to (m2);
  \path (m1) ++(1pt, -6pt) node (m1) {};
  \path (md1) ++(1pt, 0) node (md1) {};
  \draw[-Latex] (m1) to (md1.center) to[in=270, out=270, looseness=.6] (md2.center) to (m2);
\end{tikzpicture}$

\vspace{2.5\baselineskip}}}%
\hfill%
\subcaptionbox{Case L2 with $i'' \neq i'$ and $j'' = j'$.\label{fig:NL2a}}{\parbox{.47\textwidth}{
\centering$\strut$\\$\strut$\\$\stringstrut%
\str = %
\stringbox[1mm]{}{}{\stringidx{\lhs}}%
\stringbox[\tmpd]{}{\mathllap{\stringidx{i}\enskip}\alpha}{\stringidx{j = i'}}%
\stringbox[5mm]{}{}{\stringidx{(j' - \ell')\ \ \ }}%
\stringbox[\tmpd]{}{\mathllap{\stringidx{\ i''}\enskip}\alpha}{\stringidx{j'' = j'}}%
\stringbox[1mm]{}{}{}$

$\stringstrut%
\phantom{\str = \stringbox[1mm]{}{}{\stringidx{}}}%
\stringstrut%
\mathrlap{\stringbox[1.4mm]{}{\beta}{}}
\phantom{\stringbox[\tmpd]{}{\alpha}{\stringidx{}}}%
\tikzmark{m1}\mathllap{\stringbox[1.4mm]{}{\beta}{}}
\phantom{\stringbox[5mm]{}{}{\stringidx{}}}%
\mathrlap{\mathllap{\stringbox[0pt]{}{\gamma}{}}\stringbox[1.4mm]{}{\beta}{}}
\phantom{\stringbox[\tmpd]{}{\alpha}{\stringidx{}}}%
\tikzmark{m2}\mathllap{\stringbox[0pt]{}{\gamma}{}\stringbox[1.4mm]{}{\beta}{}}
\phantom{\stringbox[1mm]{}{}{}}
\begin{tikzpicture}[overlay, remember picture]
  \foreach \x in {1,2} {
     \path
     (m\x) ++(-2pt,0) node (m\x) {}
     (m\x) ++(-12.5mm,0) node (e\x) {} 
     (m\x) ++(0,-10pt) node (md\x) {}
     (e\x) ++(0,-10pt) node (ed\x) {};
  }
  \draw[-Latex] (e1) to (ed1.center) to[in=270, out=270, looseness=.6] (md1.center) to (m1);
  \draw[-Latex] (e2) to (ed2.center) to[in=270, out=270, looseness=.6] (md2.center) to (m2);
  \path (m1) ++(1pt, -6pt) node (m1) {};
  \path (md1) ++(1pt, 0) node (md1) {};
  \draw[-Latex] (m1) to (md1.center) to[in=270, out=270, looseness=.6] (md2.center) to (m2);
\end{tikzpicture}$

\vspace{2.5\baselineskip}}}

\subcaptionbox{Case L2 with $i'' = i'$ and $j'' = j'$.\label{fig:NL2b}}{%
\newcommand{\dashline}[1][-4.5cm]{\begin{tikzpicture}[overlay, remember picture]
	\draw (0,8pt) edge[dashed] ++(0, #1);
\end{tikzpicture}}%
\parbox{\textwidth}{%
\flushleft$\strut$\\$\strut$\\$\stringstrut%
\str = %
\mathrlap{\stringbox[6cm]{}{}{}}
\phantom{\stringbox{}{}{}}%
\mathllap{\stringidx{\lhs}}%
\dashline%
\mathrlap{\phantom{\stringbox[6.5mm]{}{\mathclap{\mu}}{}}%
\dashline%
\mathllap{\stringidx{j' - \ell'}}}%
\phantom{\stringbox[5mm]{}{\mathclap{\textnormal{suf}(\mu)}}{}}%
\phantom{\stringbox[6.5mm]{}{\mathclap{\mu}}{}}%
\phantom{\stringbox[6.5mm]{}{\mathclap{\mu}}{}}%
\mathllap{\stringidx{i}}\tikzmark{sh1}\dashline
\phantom{\stringbox[6.5mm]{}{\mathclap{\mu}}{}}%
\mathllap{\stringidx{j = i'' = i'}}\tikzmark{th1}\tikzmark{sh2}\dashline[-4cm]
\phantom{\stringbox[6.5mm]{}{\mathclap{\mu}}{}}%
\mathllap{\stringidx{j' = j''}}\tikzmark{th2}\dashline[-1.7cm]%
$

\vspace{1\baselineskip}

$\stringstrut%
\phantom{\str = \stringbox{}{}{\stringidx{\lhs}}%
\stringbox[6.5mm]{}{\mathclap{\mu}}{}}%
\mathrlap{\stringbox[25mm]{}{\phantom{\alpha}\mathrlap{\alpha}}{}}%
\phantom{\stringbox[5mm]{}{\mathclap{\textnormal{suf}(\mu)}}{}%
\stringbox[6.5mm]{}{\mathclap{\mu}}{}%
\stringbox[6.5mm]{}{\mathclap{\mu}}{}%
\stringbox[6.5mm]{}{\mathclap{\mu}}{}%
\stringbox{}{}{}}%
$

$\tikzmark{tv2}\stringstrut%
\phantom{\str = \stringbox{}{}{\stringidx{\lhs}}%
\stringbox[6.5mm]{}{\mathclap{\mu}}{}}%
\mathrlap{\stringbox[6.5mm]{}{\mathclap{\gamma}}{}\stringbox[18.05mm]{}{\beta}{}%
\smash{\begin{rcases*}\strut\\\strut\\\strut\end{rcases*}%
\parbox{4cm}{\centering
right occurrence $\str(j' - \ell'..j']$ of $\alpha$:
prefixes $\str[1..i'']$ and $\str[1..j'']$
share the suffix $\gamma\beta (= \alpha)$}}%
}%
\phantom{\stringbox[6.5mm]{}{\mathclap{\mu}}{}%
\stringbox[5mm]{}{\mathclap{\textnormal{suf}(\mu)}}{}%
\stringbox[6.5mm]{}{\mathclap{\mu}}{}%
\stringbox[6.5mm]{}{\mathclap{\mu}}{}%
\stringbox{}{}{}}%
$

$\tikzmark{sv2}\stringstrut%
\phantom{\str = \stringbox{}{}{\stringidx{\lhs}}}%
\mathrlap{\stringbox[6.5mm]{}{\mathclap{\gamma}}{}\stringbox[18.05mm]{}{\beta}{}}%
\phantom{\stringbox[6.5mm]{}{\mathclap{\mu}}{}%
\stringbox[5mm]{}{\mathclap{\textnormal{suf}(\mu)}}{}%
\stringbox[6.5mm]{}{\mathclap{\mu}}{}%
\stringbox[6.5mm]{}{\mathclap{\mu}}{}%
\stringbox[6.5mm]{}{\mathclap{\mu}}{}%
\stringbox{}{}{}}%
$

\vspace{2\baselineskip}

$\stringstrut%
\phantom{\str = \stringbox{}{}{\stringidx{\lhs}}}%
\mathrlap{\stringbox[25mm]{}{\alpha}{}}%
\phantom{\stringbox[6.5mm]{}{\mathclap{\mu}}{}%
\stringbox[5mm]{}{\mathclap{\textnormal{suf}(\mu)}}{}%
\stringbox[6.5mm]{}{\mathclap{\mu}}{}%
\stringbox[6.5mm]{}{\mathclap{\mu}}{}%
\stringbox[6.5mm]{}{\mathclap{\mu}}{}%
\stringbox{}{}{}}%
$

$\tikzmark{tv1}\stringstrut%
\phantom{\str = \stringbox{}{}{\stringidx{\lhs}}%
\stringbox[6.5mm]{}{\mathclap{\mu}}{}}%
\mathrlap{\stringbox[18.05mm]{}{\beta}{}%
\smash{\begin{rcases*}\strut\\\strut\\\strut\end{rcases*}%
\parbox{4cm}{\centering
left occurrence\\$\str(\lhs..i']$ of $\alpha$:\\
prefixes $\str[1..i]$ and $\str[1..j]$\\
share the suffix $\beta$}}%
}%
\phantom{\stringbox[6.5mm]{}{\mathclap{\mu}}{}%
\stringbox[5mm]{}{\mathclap{\textnormal{suf}(\mu)}}{}%
\stringbox[6.5mm]{}{\mathclap{\mu}}{}%
\stringbox[6.5mm]{}{\mathclap{\mu}}{}%
\stringbox{}{}{}}%
$

$\tikzmark{sv1}\stringstrut%
\phantom{\str = \stringbox{}{}{\stringidx{\lhs}}}%
\mathrlap{\stringbox[18.05mm]{}{\beta}{}}%
\phantom{\stringbox[6.5mm]{}{\mathclap{\mu}}{}%
\stringbox[5mm]{}{\mathclap{\textnormal{suf}(\mu)}}{}%
\stringbox[6.5mm]{}{\mathclap{\mu}}{}%
\stringbox[6.5mm]{}{\mathclap{\mu}}{}%
\stringbox[6.5mm]{}{\mathclap{\mu}}{}%
\stringbox{}{}{}}%
$

\begin{tikzpicture}[overlay, remember picture]
	\foreach \x in {1,2} { \foreach \y in {s,t} { \foreach \z in {h,v} {
	\path (\y\z\x) ++(-2pt,-3pt) node (\y\z\x) {};
	}}}
	\foreach \x in {1,2} {
		\draw[-Latex] (sh\x |- sv\x) to[in=270, out=270, looseness=.7] (th\x |- sv\x) to (th\x |- tv\x);
	}
\end{tikzpicture}

\vspace{.5\baselineskip}

}}
\caption{Additional drawings for \cref{sec:llce}, Cases L1 and L2.}
\end{figure}

\clearpage

\end{document}